# Enhanced activity in layered-metal-oxide-based oxygen evolution catalysts by layer-by-layer modulation of metal ion identity.

Ran Ding,[†,§,1] Daniel Maldonado-Lopez,[§,2] Jacob E. Henebry,[1] Jose Mendoza-Cortes,[2,3,*] Michael J. Zdilla[1,*]

[1]Deparment of Chemistry, Temple University, 1901 N. 13th St., Philadelphia, PA

[2]Department of Chemical Engineering & Materials Science, Michigan State University, East Lansing, MI 48824, USA

[3]Department of Physics & Astronomy, Michigan State University, East Lansing, Michigan 48824, USA

*Water Oxidation, Water Splitting, Layered Materials, Electrocatalysis, Heterogenous Catalysis*

**ABSTRACT:** Few-layered potassium nickel and cobalt oxides show drastic differences in catalytic activity based on metal ion preorganization. Uniform compositions [$(CoO_2/K)_6$ or $(NiO_2/K)_6$] show limited activity, while homogenously-mixed-metal cobalt/nickel oxides [$(Co_nNi_{(1-n)}O_2/K)_6$] display moderate improvement. However, a layer-by-layer arrangement of cobalt and nickel oxide sheets [e.g., $(CoO_2/K/NiO_2/K)$] provides superior catalytic performance, reducing the oxygen evolution overpotential by more than 400 mV. Density functional theory simulations provide an illustration of the electronic properties (density of states and localization of orbitals) that promote catalysis in the layer-segregated materials over those of homogeneous composition. This study reveals that atomic preorganization of metal ions within layered catalysts plays a more crucial role than overall metal composition in enhancing catalytic efficiency for oxygen evolution.

## INTRODUCTION

The pursuit of solar water splitting as an alternative to environmentally harmful burning of fossil fuels has led to extensive research on heterogeneous catalysts for water oxidation: the most challenging half of the water splitting reaction. The generation of reducing equivalents, protons, and $O_2$ from water is challenging due to the stability of water, and the common formation of reactive oxygen species as undesirable partially oxidized products.[1] The development of catalysts that perform the full four-electron oxidation of two water molecules without deleterious reactive oxygen species formation is of central importance. Although iridium and ruthenium oxides have served as benchmark oxygen evolution reaction (OER) catalysts, their high cost limits their applicability. Recently, layered materials of first-row transition metals, including nickel and cobalt, have emerged as promising, cost-effective alternatives. However, most studies on these catalysts focus on metal composition rather than atomic-scale arrangement, leaving the effect of metal ion organization on catalytic efficiency largely unexplored.

The layered double hydroxides are mixed-metal layered-catalytic materials that constitute a particularly active class of catalysts that have received much recent attention for their ability to catalyze the oxygen evolution reaction (OER) at low overpotential.[2] Other related cobalt-containing mixed-metal oxides, hydroxides, and alkoxides [3] have also shown excellent catalytic properties. Studies on layered nickel and cobalt oxides also demonstrate promise for these metal oxides in water oxidation.[4, 5] Work from our group and others' have explored the layered manganese oxide birnessite in great detail, and uncovered details about the role of interior cation identity,[6, 7-9, 10, 11] oxidation state,[8, 12, 13] solvent frustration,[7, 14, 15] and defect density[12, 13] on the activity of these catalysts. Doping of other transition metals into birnessite[10, 11, 16] has given rise to the most substantial improvements in catalysis. While birnessite traditionally requires a very large overpotential of 700 mV or more to catalyze OER (at 10 mA), simple protocols to include defects and dopants in birnessite have brought the potential to less than 400 mV, making these modified birnessites superior to any class of pure manganese-based heterogeneous OER catalyst. Use of elemental mixtures in heterogeneous catalysis in general has been a successful approach to improving OER activity,[6, 7, 8, 11] primarily due to alterations in the density of states (DOS) imposed by the dopant orbitals, which can improve hole migration, substrate binding, and transition state stabilization. However, in most studies, the dopants are distributed randomly throughout the structure with little attention paid to how the *organization* of heterometallic structures affects catalysis.

A recent study from one of our groups demonstrated a remarkable dependence of OER on the distribution of catalytically important $Mn^{III}$ in few-layer birnessites. Birnessite is a hydrated, layered manganese oxide with general formula $K_xMnO_2(H_2O)_y$. The $MnO_2$ atomically thin layers are mostly $Mn^{IV}$-based, but contain some amount (typically less than half) of $Mn^{III}$. These $Mn^{III}$ centers impart a negative charge to the sheet, which is balanced by hydrated interlayer potassium ions. While $Mn^{III}$ has long been known to promote OER in manganese-based catalysts,[17] a theoretical study from Perdew and collaborators predicted that $Mn^{III}$ abundance was *not* the main key to catalytic activity in birnessite, but rather the *organization* of $Mn^{III}$-rich layers proximal to $Mn^{III}$-poor layers. These "potential steps" between electron-rich and electron-poor sheets were predicted to facilitate electron transport across layers by positioning the small polaronic $e_g^1$ state at the valence band maximum (VBM) of the $Mn^{III}$-rich layer at similar energy to the conduction band minimum (CBM) of the adjacent $Mn^{III}$-poor layer, resulting in enhanced electron transfer. We showed in this work that few-layer catalysts with such alternating layers of $Mn^{III}$-rich and -poor layers were superior catalysts to systems constructed entirely from $Mn^{III}$-rich layers.[13] Further, we confirmed the theoretically predicted changes to the DOS using scanning tunneling spectroscopy, which showed the expected repositioning of band edges of bilayers of $Mn^{III}$-rich and -poor sheets.[12]

Since DOS is dependent not only upon oxidation state of metals, but also (and even more so) by metal identity, we report here how modulation of DOS in layered materials by a more obvious approach—the alternation of metal identity in neighboring layers—would affect OER activity. We report the preparation of heterostructured few-layer nickel and cobalt oxide materials with interstitial hydrated potassium ions. This is achieved by the exfoliation of pure phases of the layered material followed by a layer-by-layer reassembly onto fluorine-doped tin oxide electrodes with interlayer hydrated potassium ions, making these materials analogous to potassium birnessite, but with layers of pure cobalt oxide (Co) or nickel oxide (Ni), or heterostructures of each. The layer-by-layer reassembly protocol permits exquisite control of the layer order, permitting an examination of the role of the position and number of the potential steps in these structures. The result shows that the heterostructures facilitate catalysis and that the activity of these ordered materials is greater than the sum of their parts.

## RESULTS AND DISCUSSION

**Synthesis and few-layer catalyst assembly**

$LiNiO_2$ and $LiCoO_2$ were prepared using published protocols.[4, 18] For control experiments, we also prepared $LiCo_xNi_{(1-x)}O_2$ samples ($x = 1/3, 1/2, 2/3$) with homogeneously distributed cobalt and nickel *within* the sheets (i.e., solid solutions).

The layer morphology is confirmed by TEM (Fig. S1-3)-S). All samples share a similar structure, characterized by PXRD (Fig. 1). The position of the (003) peak around $2\theta = 19°$ indicates $LiNiO_2$ has a larger lattice parameter $c$ than $LiCoO_2$, which is consistent with literature.[19] The elemental composition was confirmed by ICP-OES.

Using an exfoliation and reassembly approach,[12, 15, 20] we precisely controlled $MO_2$ stacking to produce few-layered catalysts with systematic layer organization. First, we exfoliated the metal oxides into single-layer nanosheets (NS) of stoichiometric $[Co(III)O_2]_n^{n-}$ and $[Ni(III)O_2]_n^{n-}$ via insertion of bulky tetrabutylammonium ($NBu_4^+$) ions into the interlayer.[15]

A layer of PEI was deposited on an FTO substrate to create a positively charged surface to which the first negative metal oxide sheet adheres. The substrate is repeatedly coated in NS followed by a rinse, and then coated with $K^+$ using a solution of KOH followed by a rinse. Repetition of this process results in stacking sequential layers of metal oxides with intervening layers of aqueous $K^+$, to re-assemble few-layer $KMO_2$ with controllable structures (Figure 2).

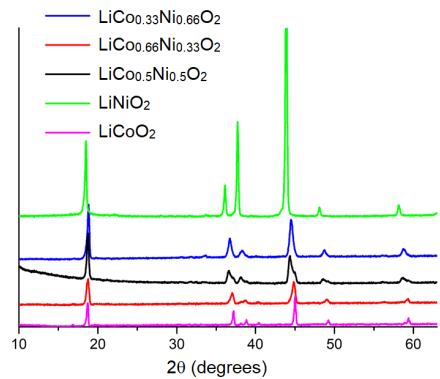

**Figure 1** Powder X-ray diffraction (PXRD) patterns of synthetic $LiCo_xNi_{1-x}O_2$ with x = 0, 0.33, 0.50, 0.66, 1). The 003 peak at $2\theta = 19°$ corresponds to the interlayer spacing. The region between $2\theta = 37$-$40$ comprises three reflections: (101, 006, and $10\overline{2}$, which have slightly different intensities stemming from changes in atomic composition. The positions exhibit a gradual shift to higher $2\theta$ as the cobalt content increases due to the slightly small lattice constants of $LiCoO_2$.[21]

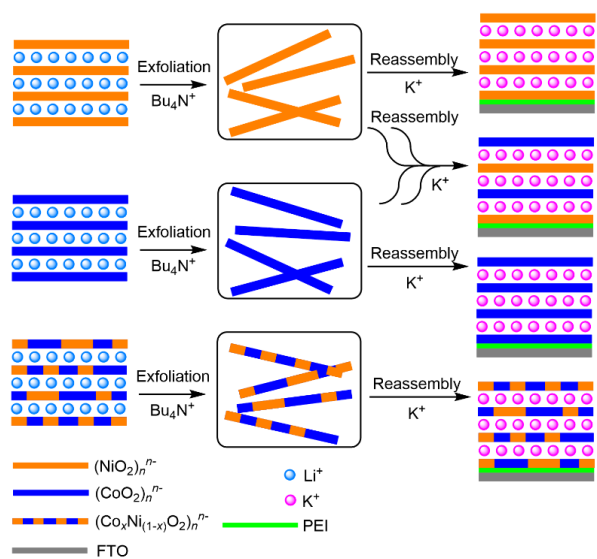

**Figure 2**. Schematic of the exfoliation and layer-by-layer reassembly process used to deposit metal oxide nanosheets on PEI-coated fluorine-doped tin oxide (FTO) substrates. Successive coatings of metal oxide layers and intercalated potassium ions yield few-layer KMO structures with precise layer order, enabling control over the atomic arrangement of Co and Ni oxide sheets.

Catalyst layer structure is denoted using, for example, $(M/K)_6$ (M = Co, Ni), indicating six layers of metal oxide with 6 layers of potassium ions; the first layer is deposited onto PEI-coated FTO. For mixed-layer materials, the leftmost indicated metal is the one stacked first against FTO. Samples were prepared by six sequential coating cycles. $(Co/K/Ni/K)_3$ was synthesized by stacking pure cobalt oxide layers with pure nickel oxide layers alternatively three times each, with $K^+$ as intercalated cations. The few-layer material $(Co_{0.5}Ni_{0.5}/K)_6$, as another example, was prepared by repeatedly stacking one layer of mixed/doped $[Co_{0.5}Ni_{0.5}O_2]^-$ with another layer of $K^+$ six times. The elemental composition and the fidelity of the stacking process is confirmed by 2p Co and Ni XPS analysis, which show the corresponding signals for cobalt and/or nickel depending on which layers have been added.

**Electrocatalysis**

Water oxidation was carried out in 1M KOH and corrected to RHE. For $[Co_{0.5}Ni_{0.5}O_2]^-$, the overpotential at the current density of 5mA/cm$^2$ is ~650 mV. As shown in Figure 3, the homogeneously alloyed structure $(Co_{0.5}Ni_{0.5}/K)_6$ exhibits improved catalytic activity compared to pure cobalt oxide $(Co/K)_6$ and pure nickel oxide $(Ni/K)_6$. Notably, the layer-by-layer arrangement $(Co/K/Ni/K)_3$, with alternating Co and Ni oxide layers, achieves an exceptional OER overpotential reduction to ~460 mV, underscoring the impact of atomic-scale layer modulation on catalytic performance. Considering both $(Co_{0.5}Ni_{0.5}/K)_6$ and $(Co/K/Ni/K)_3$ possess the same Co/Ni ratio (1:1) and the same metal oxide layer number, the improvement of the catalytic performance is attributed to the unique organized structure of the latter material. This structural feature alters the electronic structure of the material and thus improves its catalytic activity.

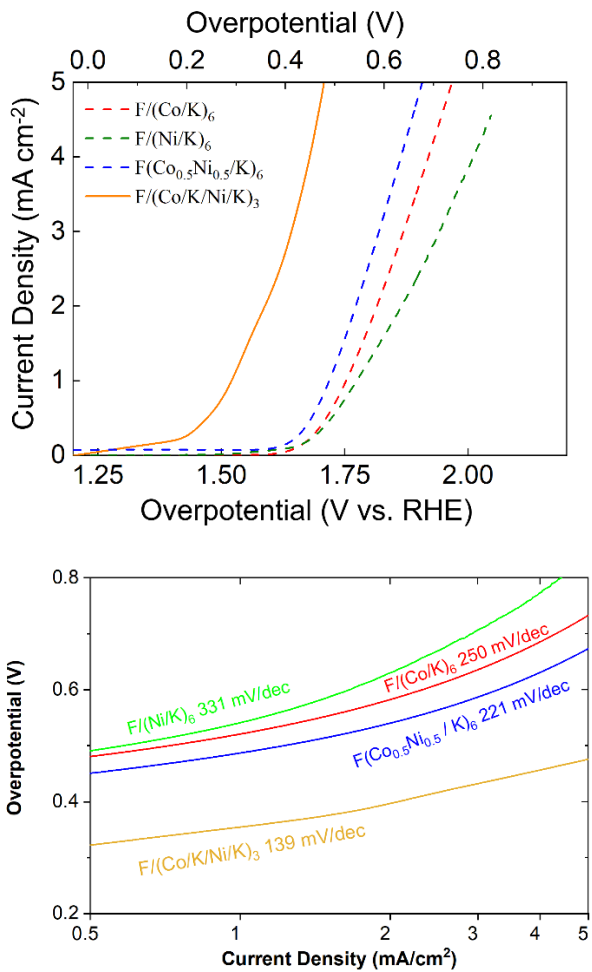

**Figure 3.** Top: Linear sweep voltammetry of few layer materials $(Co/K)_6$, $(Ni/K)_6$, $(Co_{0.5}Ni_{0.5}/K)_6$ and $(Co/K/Ni/K)_3$. All samples are made by six metal oxide layers and "F" represents FTO glass. Bottom: Tafel slopes derived from the LSV curves in the top.

In addition to the 1:1 Co:Ni ratio experiment discussed above, we have also applied this strategy to mixed metal oxides with 1:2 metal ratios. As shown in Figure 4, when the ratio of Co to Ni is 1:2, $(Co/K/Ni/K/Ni/K)_2$ gives an overpotential of ~500 mV at a current density of 10 mA/cm$^2$, while the overpotential of homogeneous $(Co_{0.33}Ni_{0.66}/K)_6$ is much higher: ~740 mV. When the Co/Ni ratio is 2:1 in $(Ni/K/Co/K/Co/K)_2$ the material shows the best catalytic performance and the overpotential is ~370 mV. By contrast, homogenous $(Co_{0.66}Ni_{0.33}/K)_6$ gives a much higher overpotential of ~600 mV. Despite some expected variability sample-to-sample, these trends are robust across multiple trials and multiple samples (Figure 5, S4). In Figure 5, we summarize the overpotential results with standard deviation bars. We can clearly see that mixed metal oxides with alternatively distributed cobalt and nickel (circles) perform better than corresponding samples with homogeneous distribution (squares). The chronoamperometry curves shown in Fig. S5 demonstrate the alternatively stacked sample could be kept active for more than 1 hour at an overpotential of less than 370 mV.

Tafel analysis (Fig. 3, 4) reveals some noteworthy trends across these materials. Tafel slope is influenced by kinetic barriers (in this case, electron transport barriers), while the vertical position is related to the number of charge carriers. While these values may be dependent on other impedance factors (such as chemical kinetics, mass transport, electrode surface area,),[22] these latter properties are well-controlled for in our use of analogous electrodes, surface area normalization, and the use of chemically analogous catalysts. Thus differences may be attributed to the internal electron transport barriers. Across these materials generally, as the overpotential increases, the Tafel slope also increases suggesting changes in activation energy are responsible for catalytic activity. The only exception is between the $F/(Co/K/Ni/K/Ni/K)_2$ and the $F/(Co_{0.66}Ni_{0.33}/K)_6$ samples (Fig. 4), which have similar Tafel slopes, but different vertical positions. The lower vertical position of the modulated sample $F/(Co/K/Ni/K/Ni/K)_2$ is consistent with the general observation that the modulation of DOS in adjacent layers provides more facile transport of electrons across the layers.

Elemental analysis (XPS) following electrocatalysis confirms that the atomic composition is mostly retained during and after OER, though some signal loss is observed upon catalyst death. The XPS of the most active mixed material shows that during catalysis, cobalt is more prone to leaching than Ni, as the Co signal has decreased by about 50% post-mortem (Fig 7A), while the Ni signals have remained at the same intensity (Fig. S7B). In few-layered catalyst of nickel- or cobalt-only layers, the ions appear to leech equivalently (about 40% post mortem). With such thin-layered materials, structural analysis of the nanolayered substrate on FTA following electrocatalysis was not practical. However, even if structural rearrangement does occur during catalysis, the results show that the periodic modulation of metal ion identity nevertheless improves catalytic performance.

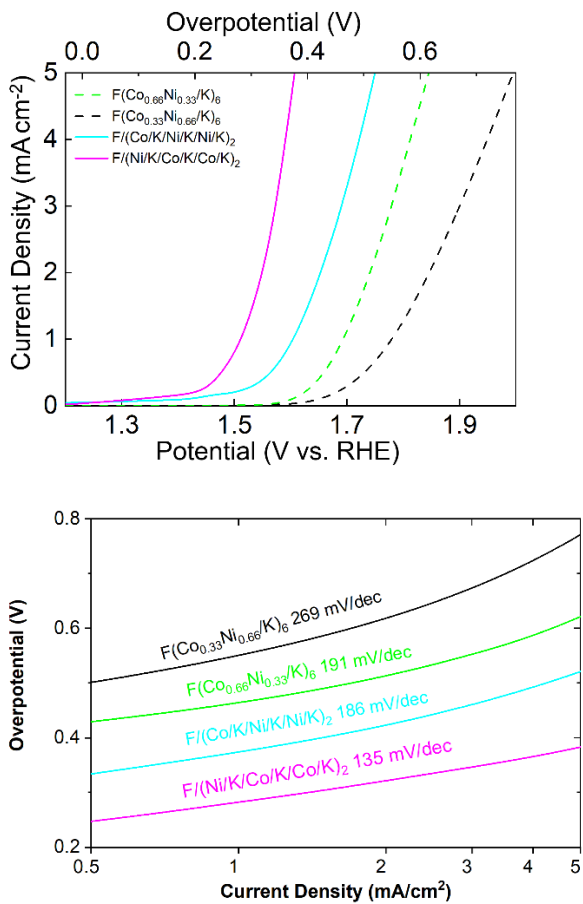

**Figure 4.** Top: Linear sweeping voltammetry of few layer materials (Co/K/Ni/K/Ni/K)$_2$, (Ni/K/Co/K/Co/K)$_2$, (Co$_{0.33}$Ni$_{0.66}$/K)$_6$ and (Co$_{0.66}$Ni$_{0.33}$/K)$_6$. All samples are made by six metal oxide layers and "F" represents FTO glass. Bottom: Tafel slopes derived from the LSV curves in the top.

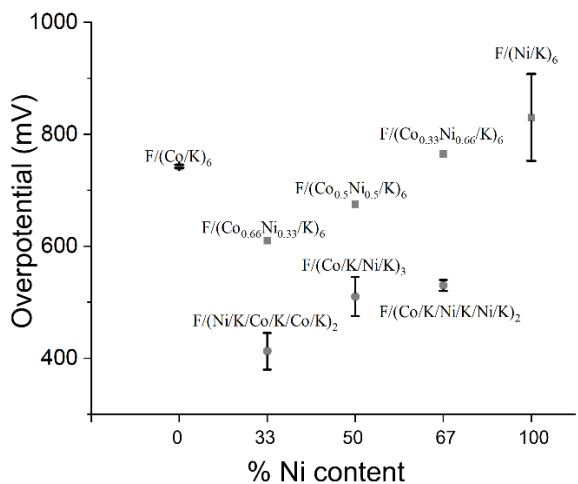

**Fig. 5** The overpotential of few layer materials at 5 mA/cm$^2$ with error bars.

While it is established from these experiments that alternating Co/K/Ni/K layers provide the best overpotentials, the question of whether the identity of the initial layer is important is of interest. We prepared a separate batch of catalysts wherein we compared six-layer alternating systems with nickel first (i.e., F/(Ni/K/Co/K)$_6$) to those with cobalt first (F/(Co/K/Ni/K), and compared these to the pure phases F/(Co/K)$_6$ and F/(Ni/K)$_6$. In both cases, the compositionally modulated heterolayered materials were superior to the compositionally uniform materials, but we found that the stacking of cobalt layers first in F/(Co/K/Ni/K) improved the overpotential by 100 mV more than when nickel was stacked first in F/(Ni/K/Co/K)$_6$, though stacking of nickel first in a mixed catalyst was better than the all-nickel catalyst (See Figure S6).

**Computational studies**

We used first-principles unrestricted hybrid Density Functional Theory with Grimme-D3 dispersion corrections (HSE06-UDFT-D3, or DFT for simplicity) to compare the electronic properties of mixed metal oxide vs. pure metal oxide heterostructures. The HSE06 hybrid functional was chosen due to its effectiveness in accurately modeling electronic properties in layered oxides, and the Grimme-D3 dispersion correction addresses interlayer interactions critical to replicating the experimental conditions of layer-stacked material. In Figure 6, we show the total band structures and density of states (DOS) of the optimized *bulk* materials. We find that different-layer (Co/K/Ni/K)$_\infty$ and same-layer (Co$_{0.5}$Ni$_{0.5}$/K)$_\infty$ both present intermediate energy levels in the 1 to 2.5 eV range (coming mainly from Ni contributions) and in the 4.5 to 6 eV range (coming mainly from Co contributions), which are not present in the pure layered materials. The formation of inter-gap states is an excellent preliminary indication that K-intercalated mixed metal oxide heterostructures allow better electron conduction throughout the structure (Figure 6).

Due to these intermediate energy levels, the bulk materials show promising behavior for increased catalytic activity in the mixed-transition-metal structures compared to the pure compounds. However, the bulk electronic properties do not replicate the trends observed in the experimental overpotentials of the few-layer materials. Moreover, the gaps in these bulk materials are too large for efficient electron conduction. Therefore, we next generated and analyzed the few-layer heterostructures. To accomplish this, we generated slabs, emulating the experimental synthetic structures. These structures are periodic in the *x* and *y* direction, simulating infinitely large sheets. The KMO$_2$ formula unit is repeated 6 times in the *z* direction. In Figure 7, we present the spin-polarized DOS for these few-layer compounds; in this figure; spin-up electrons are plotted to the right side and spin-down electrons to the left for each DOS plot. From the plots in Figure 7, it is evident that few-layered structures have better electron-transfer properties than bulk structures, as the gaps observed in bulk DOS close for almost all slab compounds, allowing better conduction throughout the structure.

Water splitting is thermodynamically possible at a voltage of 1.23 V vs. SHE.[23] This energy corresponds to -5.67 eV w.r.t. vacuum, and is abbreviated the standard oxygen electrode (SOE) level. Therefore, it is important that catalysts for oxygen evolution provide enough electronic states in the SOE level, facilitating the generation of charge carriers with sufficient energy to participate in the reaction. This can be analyzed through the DOS plots in Figure 7. It is important to note that the plots were aligned with respect to vacuum to make direct comparisons between structures' DOS profiles. Furthermore, the width of the DOS (x-axis) was normalized to the largest DOS peak of each individual structure in the -10 to 0 eV range. In these plots, the dotted black line indicates the Fermi level and the dotted green line represents the SOE level (-5.67 eV w.r.t. vacuum). Furthermore, the filled grey rectangles indicate the SOE energy level plus the experimental overpotential (SOE+OP) of each corresponding structure (Figure 3). The SOE+OP region is an important descriptor of these materials' catalytic performance, as it indicates the electronic states that are likely being utilized during oxygen evolution catalysis.

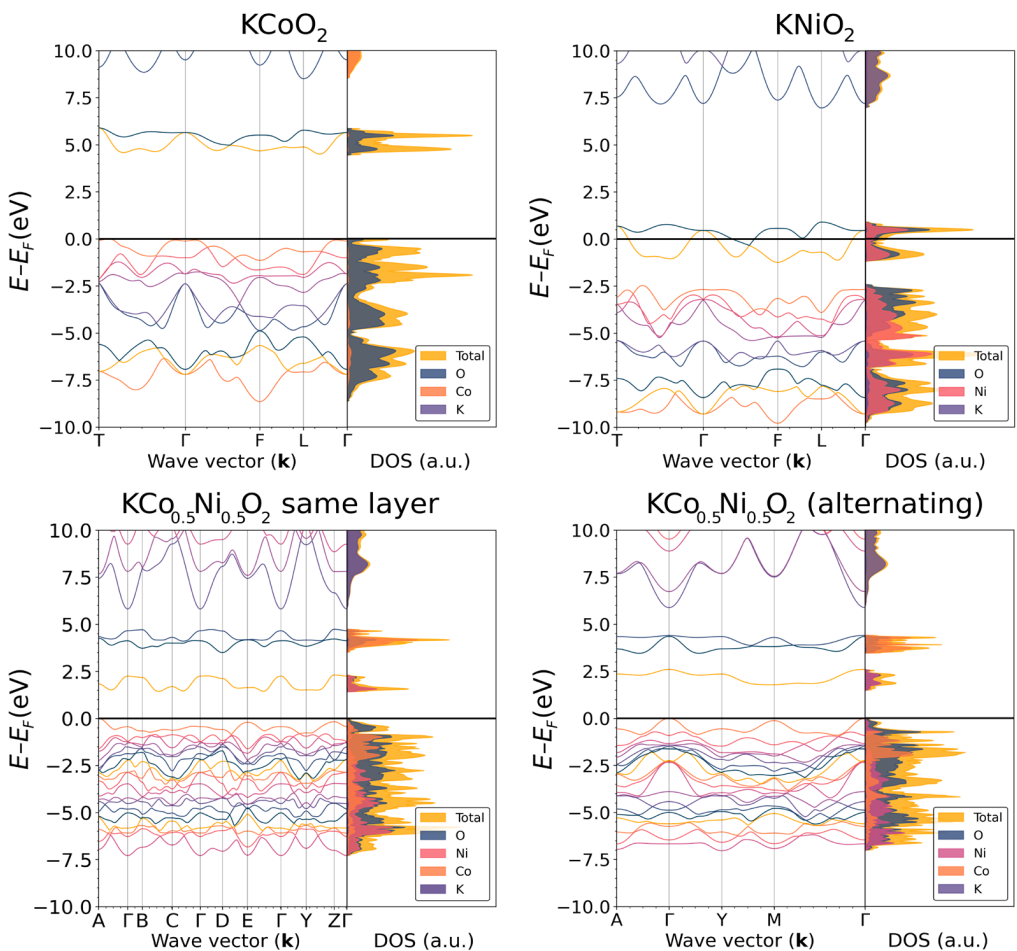

**Figure 6**. Band structures and total density of states for bulk $KCoO_2$, $KNiO_2$, same-layer $KCo_{0.5}Ni_{0.5}O_2$, and different-layer $KCo_{0.5}Ni_{0.5}O_2$ birnessites. The y-axis is centered at the Fermi level.

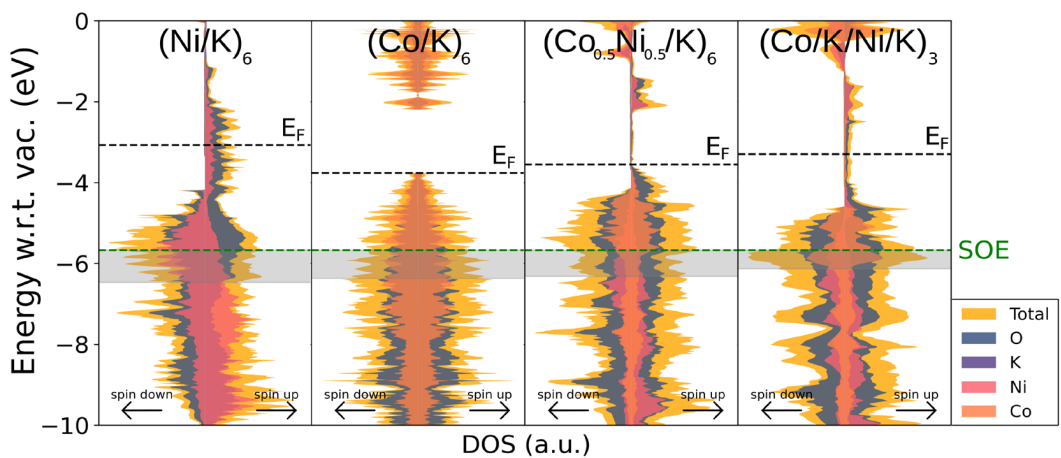

**Figure 7.** Electronic density of states of the optimized potassium-intercalated transition metal slabs. Deconvolutions for contributions from each atomic species are included. Each plot was normalized with respect to its maximum peak in the -10 eV to 0 eV energy range (w.r.t. vacuum). The dotted green line indicates the potential for oxygen evolution to occur (SOE level, -5.67 eV w.r.t. vacuum). The filled rectangles represent the "SOE plus experimental overpotential" range. The dotted black line indicates the Fermi level. Spin-up electrons are plotted to the right and spin-down electrons to the left of each DOS plot.

The plots in Figure 7 are ordered from highest to lowest overpotential (i.e., from worst to best water oxidation properties). It can be observed that $(Ni/K)_6$ and $(Co/K)_6$ present similar DOS profiles in the SOE+OP region. However, $(Ni/K)_6$ presents an uneven distribution of α (spin-up) and β (spin-down) electrons, with a smaller amount of spin-up states. This effect does not occur in $(Co/K)_6$ which presents a symmetric distribution of α and β electronic states, likely contributing more total states around the critical SOE energy level, providing more reactive carriers and resulting in a lower overpotential compared to the nickel slab. On the other hand, the mixed-composition-layer $(Co_{0.5}Ni_{0.5}/K)_6$ heterostructure presents a higher α-electron peak at the SOE level, and a relatively symmetric α and β electron distribution. Furthermore, all gaps close completely in this structure, resulting in better electron transport than the pure metal oxide slabs. Finally, the alternately stacked $(Co/K/Ni/K)_3$ heterostructure presents the largest DOS peaks for both alpha and beta electrons exactly at the SOE+OP region, resulting in the highest number of available OER-catalytic states and lowest overpotential of all calculated structures.

To further explore the catalytic properties in the alternately-stacked heterostructure, we plot the layer-resolved DOS, shown in Figure 8. Here, we observe that the outermost layers provide electron density around the Fermi level, in the -4 to -2 eV energy range. The appearance of these surface states is crucial for electron conduction in the structure, since the inner layers do not close the material's band gap completely. Through this mechanism, electron conduction is facilitated at the surface. On the other hand, inner $NiO_2$ layers provide states directly above and below the -4 to -2 eV range, facilitating electron conduction for different energy ranges. Oxygen atoms in the $NiO_2$ layers have an important contribution to the DOS in the SOE+OP range (vide infra). Finally, inner $CoO_2$ layers have strong contributions to the electron density at the SOE+OP range, especially stemming from Co atoms.

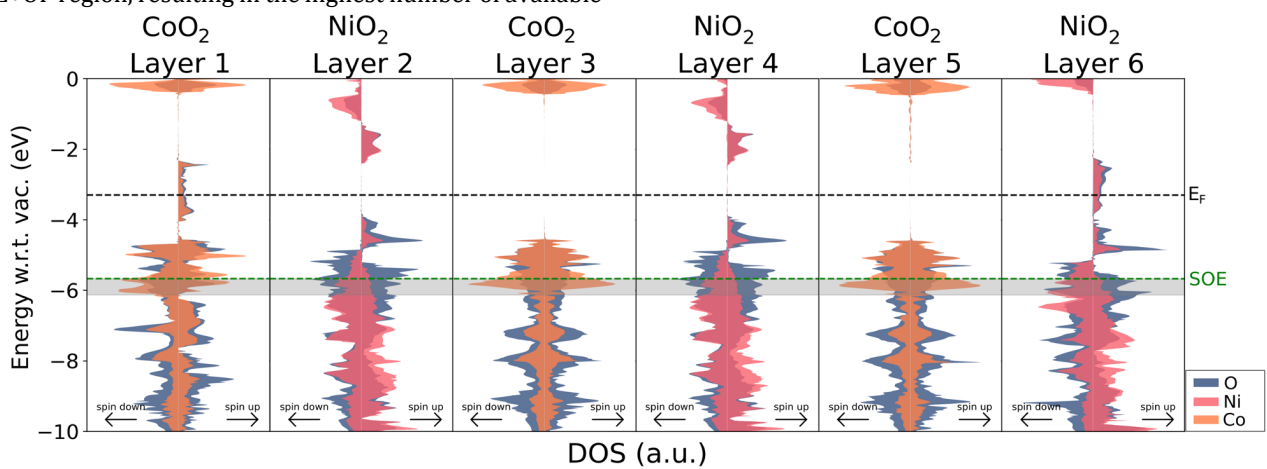

**Figure 8.** Layer-resolved DOS for the alternately-stacked $(Co/K/Ni/K)_3$ heterostructure. Layers are labeled 1-6, starting with $CoO_2$ which is completely exposed to vacuum and ending in $NiO_2$ which has a potassium layer before the vacuum. The plots were normalized to 1/5 of the maximum total DOS peak in the -10 to 0 eV range. The dotted green line indicates the potential for oxygen evolution to occur (SOE level, or -5.67 eV w.r.t. vacuum). The filled rectangles represent the "SOE plus experimental overpotential" range. The dotted black line indicates the Fermi level. Spin-up electrons are plotted to the right and spin-down electrons to the left of each DOS plot.

As previously mentioned, oxygen in $(Co/K/Ni/K)_3$ is one of the main contributors to the electron density at the SOE+OP region. This effect does not occur as drastically in the pure metal oxide structures, indicating that oxygen atoms are likely important contributors to the enhanced catalytic properties of the alternately-stacked heterostructure. To investigate this effect, we plot the spin density of the alternately stacked structure, shown in Figure 9a. The spin density plot reveals a redistribution of electrons in the $(Co/K/Ni/K)_3$ slab through the presence of pure nickel oxide layers, separated by cobalt oxide layers. In particular, we observe that the oxygens directly bonded to nickel atoms present open-shell triplet character. This triplet character can be observed in Figure 9b, which is a close-up of the outermost $NiO_2$ layer. Triplet states are generally unstable for covalent bonds, but they are highly energetic and have been linked to catalytic activity during photosynthesis,[23] and in transition metal catalysts.[24] This re-distribution of electron density is expected to promote generation of reactive oxygen species, which likely serve as catalytic sites. Finally, in Figure 9c, we plot the orbital-resolved DOS for the alternately stacked $(Co/K/Ni/K)_3$ heterostructure. Here, we observe that the main contributions to the DOS come from oxygen p-orbitals, cobalt d-orbitals, and nickel d-orbitals, with oxygen providing most of the electron density in the -10 to 0 eV range.

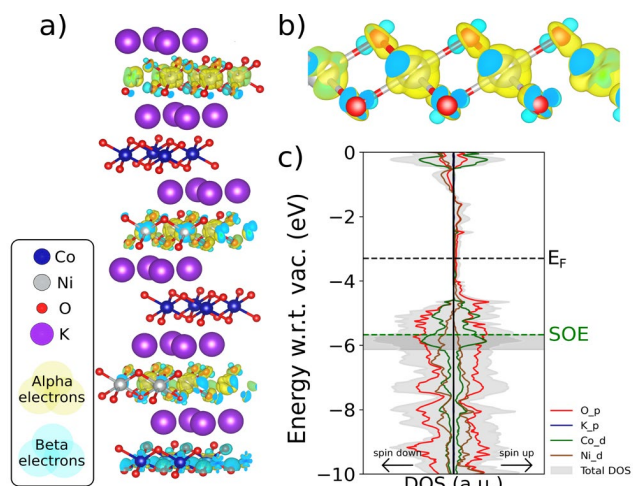

**Figure 9.** (Co/K/Ni/K)$_3$ heterostructure electronic properties. a) Spin density plot for the entire structure. b) Spin density plot for the outermost NiO$_2$ layer. c) Orbital-resolved density of states. The dotted black line indicates the Fermi level, the dotted green line indicates the standard oxygen electrode level, and the filled rectangle represents the "SOE plus experimental overpotential" range. Spin-up electrons are plotted to the right and spin-down electrons to the left of the DOS plot.

In summary, calculations indicate that slab heterostructures present more prominent gap states than bulk materials, and suggest that few-layer samples may experience a greater enhancement from these effects than bulk materials. Out of the four calculated slabs, we find that the alternately stacked (Co/K/Ni/K)$_3$ structure presents the best properties for oxygen evolution, since it provides (1) the largest amount of states at the SOE+OP energy range, (2) gap closure providing better electron transport properties, (3) crucial states around the Fermi level localized at the surface of the slab facilitating electron transport to the center of the slab, (4) inner-layer states spanning a wide range of energy levels near the SOE level, and (5) oxygen atoms in NiO$_2$ layers presenting open-shell triplet character, expected to promote OER activity.

## CONCLUSIONS

This study reveals that controlling the atomic-scale distribution of metal ions within layered transition metal oxides is a more effective approach to enhancing OER activity than altering metal composition alone. Specifically, stacking discrete Co and Ni oxide layers, rather than forming homogeneously mixed layers, reduces the OER overpotential significantly.. The superiority of preorganized structures with segregated metal identity in adjacent layers over solid solutions demonstrates that atomic organization is more important to catalytic activity than composition; Samples with Co and Ni stacked separately possess better OER performance that those with Co and Ni homogeneously distributed within the layers. The improvement is attributed to potential steps created by alternatively stacked Co and Ni oxide layers, which alters the electronic structure and facilitates electron transfer within the sample. This work, along with previous demonstration of transition metal oxides with specific distribution of charge for better OER activities, has shown the promise of this selective few-layer stacking methodology to improve the catalytic activity of a wide range of layered materials for OER, and possibly for heterogeneous catalysis of other important redox reactions. These findings underscore the potential of layer-by-layer assembly as a design strategy for electrocatalysts in OER and other redox reactions, encouraging future studies to explore atomic-scale organization for catalytic enhancement..

## EXPERIMENTAL

### General

All chemicals were purchased from chemical vendors and used without further purification except where otherwise noted. LiCoO$_2$ was purchased from Thermo Scientific, or synthesized (Vide infra). LiNiO$_2$ was purchased from Sigma Aldrich or synthesized (Vide infra). PXRD was performed on a Bruker D8 ADVANCE diffractometer using copper Kα radiation from a sealed tube or a Bruker Kappa PHOTON III DUO with a Cu μS tube. The XRD data was processed using DIFFRACEVA software packages. Inductively Coupled Plasma Optical Emission Spectroscopy samples were digested in 0.5 M hydroxylamine hydrochloride and analyzed to quantify elemental compositions using a Thermo Scientific iCAP 7000 Series ICP-OES. TEM samples were prepared by depositing one drop of NS suspension sample in water (100 mg/L) on a lacey carbon copper mesh TEM grid (400 mesh, Ted Pella) and allowed to air dry. Images were collected using a JEOL JEM-1400 microscope operating at 120 kV. Electrochemical measurements were performed using a CH Instruments CHI660E electrochemical analyzer. X-ray photoelectron spectroscopy was performed on a Thermo Scientific K-Alpha+ XPS at the University of Delaware.

### Synthesis of LiCoO$_2$, LiNiO$_2$ and LiCo$_x$Ni$_y$O$_2$.

The lithium transition metal oxides were prepared using published protocol.[4, 18] In all cases, cotton pads were soaked in solutions of lithium nitrate (LiNO$_3$) and transition metal nitrate (Co(NO$_3$)$_2$, Ni(NO$_3$)$_2$), purchased from Sigma-Aldrich. These were dissolved in de-ionized water with a 1:1 molar ratio and a total salt concentration of 0.5 M. For example, 50mL solution containing 0.5 M LiNO$_3$ and 0.5 M Co(NO$_3$)$_2$ was used to synthesize LiCoO$_2$. After soaking for 3 h, the cotton was taken out and squeezed to remove excessive liquid before being placed in a furnace and heated in air at a rate of 100 °C per hour to a temperature of 400 °C. After cooling, the resulting material was gently ground by mortar and pestle and returned to the furnace to be heated at the same rate to a final temperature of 900 °C, which was maintained overnight. Subsequently, the product was taken from the furnace and cooled in air to room temperature. Finally, the product was washed with DI water, filtered and air dried. The identities of the solids were confirmed using and compared to reported PXRD.[19] From ICP-OES, the ratio of Li/Co LiCoO$_2$ is 1.006 to 1. The ratio of Li/Ni in LiNiO$_2$ is 0.984 to 1.

### Synthesis of single layer nanosheets (NS).[23]

The intercalated Li$^+$ was removed by acid exchange by stirring 0.2 g materials LiMO$_2$ (M = Co, Ni) in dilute HNO$_3$ solution (0.1 M, 10mL) for 2 days. The suspension was centrifuged at 7000 rpm/4000g and the centrifugate washed with water until pH was between 6 and 7. Then an aqueous solution of tetra-*n*-butylammonium hydroxide (TBAOH) was added to the solid to perform exfoliation. After stirring for 10 days, the TBAOH solution was removed by centrifugation at 14000 rpm/16000g for 15-30 min and the precipitate was collected and washed by resuspending in ethanol and centrifuging again at 14000 rpm/16000g for 20 min, resuspending in water and centrifuging again at 14000/16000g for an additional 30 min. The pellet was then dispersed again into water to form a colloidal solution of purified NS. The exfoliated NS supernatant was separated from incompletely dissolved colloidal particles by and additional centrifugation at 7000 rpm/4000g for 15-25min, and the NS supernatant was decanted. Note that NS will form a pellet upon centrifugation at 14000 rpm/16000g, but remain in the supernatant at 7000 rpm/4000g.

**Assembly of NS into few layer materials.**

This layer-by-layer assembly procedure was based upon the descriptions used in previous studies.[15, 20] First, an 1cm×1cm sample of fluorine-doped tin oxide (FTO) was washed with acetone, ethanol and water, and dried in air. The substrate, dry FTO, then was immersed in to a Polyethylenimine (PEI) solution (2.5g/ L) for 1 min to be coated. After rinsing with water 3 times and drying in air, the colloidal solution of NS (manganese oxide, cobalt oxide and nickel oxide) was drop cast on to the PEI-precoated FTO to form an ultra-thin NS layer. Paper towel was used to wick away the excessive solution before the film was washed by water and dried in air for 10 min. After that, a KCl solution (5 mM) was drop cast on to the film, then the excess KCl solution was wicked away with a paper towel, rinsed with water and dried. Afterwards, the procedure is repeated with alternative coating of NS followed by K$^+$. After up to 6 sheets are deposited, the materials are terminated by a K$^+$ layer for charge balance and a final water rinse. The fidelity of the deposition process is consistent with XPS analysis on substrates following deposition of Co-only, Ni-only, or mixed Co-Ni few-layer materials (Figure S7).

**Electrochemical measurements.**

Electrochemical characterization of the various layered materials was performed using linear sweep voltammetry with a sweep rate of 0.01 V/s in 1 M KOH using a Pt wire counter electrode, a saturated calomel electrode (SCE) or an Ag/AgCl electrode. The calibration of the reference electrode was checked against the redox couple of potassium ferricyanide in water. The working electrode was FTO (1 cm×1 cm) (coated with the layered material of interest). We compared the effect of unpurified KOH electrolyte to that that had been purified to remove Fe using a published procedure[25]. We noted negligible difference in the LSV, and thus conclude contaminant iron does not influence the catalysis. Unpurified commercial KOH was used for the remainder of experiments. The internal resistivity of the apparatus as measured by impedance spectroscopy, was very low (0.5 ohms) due to the high-concentration electrolyte, and as such the ∼ 5 mV ohmic drop correction as not applied. The measured potentials were converted to the RHE scale via the Nernst equation:

$$E_{RHE} = E_{SCE} + 0.059pH + E°_{ref}$$

where $E_{RHE}$ is the converted potential vs RHE, $E_{ref}$ is the experimental potential measured against reference electrode, and $E°_{ref} = 0.2412V$ for SCE at 25 °C vs. RHE. The electrocatalysis was measured using linear sweep voltammetry (LSV) with a scan rate of 10 mV/s. The activity is normalized by dividing the current by the electrode area (∼1 cm$^2$) to give currents in units of mA·cm$^{-1}$.

In comparing individual samples for OER, it is important to note that the activity of each catalyst is highly dependent upon the synthetic batch, and we observed, depending on batch, excellent or poor catalysts for both KNiO$_2$ and KCoO$_2$, due to differences in M$^{II}$ concentration and defect density (Figure S4, S6). To control for variability, the self-same synthetic batches were always compared to one another in mixed-metal and homogeneous catalysts for effective control experiments.

## COMPUTATIONAL METHODS

Computationally, the structures were initialized from bulk structures for LiCoO$_2$[49] and LiNiO$_2$,[19] To obtain the required potassium-intercalated structures, Li ions were replaced with K ions. Bulk KCoO$_2$, KNiO$_2$, same-layer KCo$_{0.5}$Ni$_{0.5}$O$_2$, and different-layer KCo$_{0.5}$Ni$_{0.5}$O$_2$ structures were generated and fully optimized in their atomic positions and cell parameters. Furthermore, after testing different spin configurations (low-spin, intermediate-spin, and high-spin) for each model, we found that all bulk compounds are most stable in their low-spin configurations (see Figure S8).

Using the optimized bulk structures as a template, the following 24 atomic-layer computational surface structures were generated: pure cobalt oxide (Co/K)$_6$, pure nickel oxide (Ni/K)$_6$, and two alternately-stacked nickel/cobalt oxides—one with Ni & Co in the same layer (Co$_{0.5}$Ni$_{0.5}$/K)$_6$, and another with Ni & Co in different layers (Co/K/Ni/K)$_3$. These structures were fully optimized in their atomic positions and cell parameters, exploiting their symmetry when possible. The computational slab models are shown in Figure 10. Surfaces were initialized from the bulk structures performing a slab cut perpendicular to the (001) plane to generate the few-layer models. We assigned a vacuum of 500 Å (CRYSTAL17's default vacuum size for slab calculations) in the [001] direction for all surface models to avoid self-interactions arising from periodic boundary conditions. The models are equivalent to the experimental structures, starting with a transition metal oxide layer and ending in a potassium layer. We did not initiate the layered structures from the FTO glass used in the experimental procedure, as it is impossible to model non-periodic glass structures at this level of theory. Several low-, intermediate- and high-spin configurations for the slabs were tested. In this case, it was found that SCF convergence criteria were only met in a reasonable number of steps (≤ 800 SCF steps) for either low or intermediate spin order, depending on the atomic configuration.

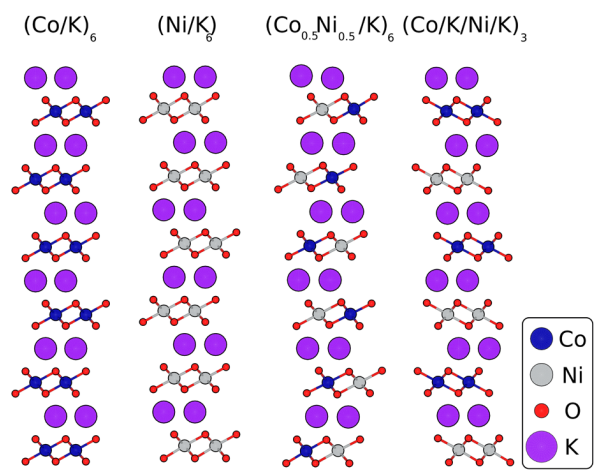

(Co/K)$_6$   (Ni/K)$_6$   (Co$_{0.5}$Ni$_{0.5}$/K)$_6$   (Co/K/Ni/K)$_3$

- Co
- Ni
- O
- K

**Figure 10.** Slab models for $(Co/K)_6$, $(Ni/K)_6$, $(Co_{0.5}Ni_{0.5}/K)_6$, and $(Co/K/Ni/K)_3$.

Geometry optimizations and electronic structures for these materials were computed using unrestricted hybrid Density Functional Theory (UDFT). Calculations were performed using the Heyd–Scuseria–Ernzerhof (HSE06) exchange-correlation functional[26] as implemented in the CRYSTAL17 code.[27] We chose HSE06 due to its high reliability in accurately predicting structural and electronic properties.[28] Furthermore, CRYSTAL17 uses Gaussian-type orbitals, which allow an efficient implementation of Post-Hartree–Fock methods. All unit cell optimizations and single point calculations were performed using triple-zeta with polarization quality (TZVP) Gaussian basis sets for K, Co, Ni, and O atoms.[29] The semiempirical Grimme-D3 dispersion correction was used to estimate Van der Waals forces.[30] The Direct Inversion of the Iterative Subspace (DIIS) convergence accelerator was used for all optimizations and single-point energy calculations.[31]

The convergence threshold for energy, forces, and electron density was $10^{-7}$ a.u. for all parameters. The reciprocal space for all the structures was sampled using a Γ-centered Monkhorst–Pack scheme with a resolution of around $2\pi \times 1/60$ Å$^{-1}$. During geometry optimization and single-point energy calculations, the SPIN and SPINLOCK keywords were used to specify the unrestricted wave functions and total spin of the transition metals. Furthermore, the ATOMSPIN keyword was used to specify unpaired electrons for the individual transition metals. High-symmetry k-points coordinates for the bulk materials' band structures were obtained using the SeeK-path software,[32] and are unique to each geometry. Density of states calculations were performed for all bands, and are shown in the -10 to 10 eV range (around the Fermi level) for bulk calculations and -10 to 0 eV range (w.r.t. vacuum) for slab calculations. Absolute band alignment was performed by running single-point calculations using the optimized structures, with ghost atoms in the vicinity of the slabs' outermost layers for a better description of the electrostatic potential in vacuum. Slab DOSs were shifted relative to the vacuum level to obtain a more accurate description of the materials' absolute band alignment. The vacuum level was defined as the asymptotic value of the plane-averaged electrostatic potential, sufficiently far away from the slab (~40Å in our calculations).

## ASSOCIATED CONTENT

**Supporting Information**. Additional electrochemical data, EM, and additional computational results. This material is available free of charge via the Internet at http://pubs.acs.org.

## AUTHOR INFORMATION

**Corresponding Author**
* mzdilla@temple.edu, jmendoza@msu.edu

**Present Addresses**
†Department of Materials Science and Engineering, North Carolina State University, 3002 Engineering Building 1, 911 Partners Way, Raleigh NC 27695

**Author Contributions**
The manuscript was written through contributions of all authors.

§These authors contributed equally.

**Funding Sources**
Department of Energy: DE-SC0012575, DE-SC0023356
National Science Foundation: CHE-2215854

## ACKNOWLEDGMENT

This work was supported as part of the Center for Complex Materials, an Energy Frontier Research Center funded by the US Department of Energy (DOE), Office of Science, Basic Energy Sciences under Award DE-SC0012575, and under an additional research grant from DOE under Award DE-SC0023356. This work was supported through computational resources and services provided by the Institute for Cyber-Enabled Research (ICER) at Michigan State University. XRD characterization was enabled by an NSF Major Research Instrumentation award CHE-2215854.

## ABBREVIATIONS

OER, Oxygen Evolution Reaction; DOS, Density of States; VBM, Valence Band Maximum; CBM, Conduction Band Minimum; TEM, Transmission Electron Microscopy; PXRD, Powder X-Ray Diffraction; ICP-OES, Inductively Coupled Plasma-Optical Emission Spectroscopy; PEI, Polyethyleneimine; FTO, Fluorine-doped Tin Oxide; TBAOH, tetrabutylammonium hydroxide; NS, nanosheet; RHE, Reversible Hydrogen Electrode; LSV, Linear Sweep Voltammetry; DFT, Density Functional Theory; UDFT, Unrestricted Density Functional Theory; SOE, Standard Oxygen Electrode; SOE + OP, Standard Oxygen Electrode + Overpotential; SCE, Saturated Calomel Electrode.

## REFERENCES

(1) Nocera, D. G. The Artificial Leaf. *Acc. Chem. Res.* **2012**, *45* (5), 767-776. DOI: 10.1021/ar2003013 (acccessed 2013/10/31).

(2) Liu, P. F.; Yang, S.; Zhang, B.; Yang, H. G. Defect-Rich Ultrathin Cobalt–Iron Layered Double Hydroxide for Electrochemical Overall Water Splitting. *ACS Applied Materials & Interfaces* **2016**, *8* (50), 34474-34481. DOI: 10.1021/acsami.6b12803. Dionigi, F.; Zeng, Z.; Sinev, I.; Merzdorf, T.; Deshpande, S.; Lopez, M. B.; Kunze, S.; Zegkinoglou, I.; Sarodnik, H.; Fan, D.; et al. In-situ structure and catalytic mechanism of NiFe and CoFe layered double hydroxides during oxygen evolution. *Nature Communications* **2020**, *11* (1), 2522. DOI: 10.1038/s41467-020-16237-1. Anantharaj, S.; Karthick, K.; Kundu, S. Evolution of layered double hydroxides (LDH) as high performance water oxidation electrocatalysts: A review with insights on structure, activity and mechanism. *Materials Today Energy* **2017**, *6*, 1-26. DOI: https://doi.org/10.1016/j.mtener.2017.07.016. Zou, X.; Goswami, A.; Asefa, T. Efficient Noble Metal-Free (Electro)Catalysis of Water and Alcohol Oxidations by Zinc-Cobalt Layered Double Hydroxide. *J. Am. Chem. Soc.* **2013**, *135* (46), 17242-17245, Article. DOI: 10.1021/ja407174u. Shao, M.; Zhang, R.; Li, Z.; Wei, M.; Evans, D. G.; Duan, X. Layered double hydroxides toward electrochemical energy storage and conversion: design, synthesis and applications. *Chem. Commun.* **2015**, *51* (88), 15880-15893, 10.1039/C5CC07296D. DOI: 10.1039/C5CC07296D.

(3) Zhu, J.; Chen, J.; Li, X.; Luo, K.; Xiong, Z.; Zhou, Z.; Zhu, W.; Luo, Z.; Huang, J.; Li, Y. Steering surface reconstruction of hybrid metal oxides for efficient oxygen evolution reaction in water splitting and zinc-air batteries. *Journal of Energy Chemistry* **2024**, *92*, 383-393. DOI: https://doi.org/10.1016/j.jechem.2024.01.020. Zhang, S.;

Huang, S.; Sun, F.; Li, Y.; Ren, L.; Xu, H.; Li, Z.; Liu, Y.; Li, W.; Chong, L.; et al. Exciting lattice oxygen of nickel–iron bi-metal alkoxide for efficient electrochemical oxygen evolution reaction. *Journal of Energy Chemistry* **2024**, *88*, 194-201. DOI: https://doi.org/10.1016/j.jechem.2023.09.013.

(4) Lu, Z.; Wang, H.; Kong, D.; Yan, K.; Hsu, P.-C.; Zheng, G.; Yao, H.; Liang, Z.; Sun, X.; Cui, Y. Electrochemical tuning of layered lithium transition metal oxides for improvement of oxygen evolution reaction. *Nature Communications* **2014**, *5* (1), 4345. DOI: 10.1038/ncomms5345.

(5) Wang, P. P.; Fu, P.; Ma, J. P.; Gao, Y. Y.; Li, Z.; Wang, H.; Fan, F. T.; Shi, J. Y.; Li, C. Ultrathin Cobalt Oxide Interlayer Facilitated Hole Storage for Sustained Water Oxidation over Composited Tantalum Nitride Photoanodes. *Acs Catalysis* **2021**, *11* (20), 12736-12744. DOI: 10.1021/acscatal.1c03298. Walton, A. S.; Fester, J.; Bajdich, M.; Arman, M. A.; Osiecki, J.; Knudsen, J.; Vojvodic, A.; Lauritsen, J. V. Interface Controlled Oxidation States in Layered Cobalt Oxide Nanoislands on Gold. *Acs Nano* **2015**, *9* (3), 2445-2453. DOI: 10.1021/acsnano.5b00158. Sun, L.; Dai, Z. F.; Zhong, L. X.; Zhao, Y. W.; Cheng, Y.; Chong, S. K.; Chen, G. J.; Yan, C. S.; Zhang, X. Y.; Tan, H. T.; et al. Lattice strain and atomic replacement of CoO6 octahedra in layered sodium cobalt oxide for boosted water oxidation electrocatalysis. *Applied Catalysis B-Environmental* **2021**, *297*. DOI: 10.1016/j.apcatb.2021.120477. Miao, X. B.; Zhou, S. M.; Wu, L.; Zhao, J. Y.; Shi, L. Spin-State Transition Enhanced Oxygen Evolving Activity in Misfit-Layered Cobalt Oxide Nanosheets. *Acs Sustainable Chemistry & Engineering* **2018**, *6* (9), 12337-12342. DOI: 10.1021/acssuschemeng.8b02804. Liu, R. C.; Liang, F. L.; Zhou, W.; Yang, Y. S.; Zhu, Z. H. Calcium-doped lanthanum nickelate layered perovskite and nickel oxide nano-hybrid for highly efficient water oxidation. *Nano Energy* **2015**, *12*, 115-122. DOI: 10.1016/j.nanoen.2014.12.025. Lin, X.; Bao, H. L.; Zheng, D. H.; Zhou, J.; Xiao, G. P.; Guan, C. Z.; Zhang, L. J.; Wang, J. Q. An Efficient Family of Misfit-Layered Calcium Cobalt Oxide Catalyst for Oxygen Evolution Reaction. *Advanced Materials Interfaces* **2018**, *5* (23). DOI: 10.1002/admi.201801281. Li, Y.; Chen, G.; Zhu, Y. P.; Hu, Z. W.; Chan, T. S.; She, S. X.; Dai, J.; Zhou, W.; Shao, Z. P. Activating Both Basal Plane and Edge Sites of Layered Cobalt Oxides for Boosted Water Oxidation. *Adv. Funct. Mater.* **2021**, *31* (38). DOI: 10.1002/adfm.202103569. Kuznetsov, D. A.; Konev, D. V.; Sokolov, S. A.; Fedyanin, I. V. Cobalt Oxide Materials for Oxygen Evolution Catalysis via Single-Source Precursor Chemistry. *Chemistry-a European Journal* **2018**, *24* (52), 13890-13896. DOI: 10.1002/chem.201802632. Hu, X. M.; Zhang, S. L.; Sun, J. W.; Yu, L.; Qian, X. Y.; Hu, R. D.; Wang, Y. N.; Zhao, H. G.; Zhu, J. W. 2D Fe-containing cobalt phosphide/cobalt oxide lateral heterostructure with enhanced activity for oxygen evolution reaction. *Nano Energy* **2019**, *56*, 109-117. DOI: 10.1016/j.nanoen.2018.11.047. Dai, J.; Zhu, Y. L.; Chen, Y. B.; Zhou, W.; Shao, Z. P. Na0.86Co0.95Fe0.05O2 Layered Oxide As Highly Efficient Water Oxidation Electrocatalyst in Alkaline Media. *Acs Applied Materials & Interfaces* **2017**, *9* (26), 21587-21592. DOI: 10.1021/acsami.7b06004. Chen, Z.; Cai, L.; Yang, X. F.; Kronawitter, C.; Guo, L. J.; Shen, S. H.; Koel, B. E. Reversible Structural Evolution of NiCoOxHy during the Oxygen Evolution Reaction and Identification of the Catalytically Active Phase. *Acs Catalysis* **2018**, *8* (2), 1238-1247. DOI: 10.1021/acscatal.7b03191. Barauskiene, I.; Valatka, E. Layered Nickel-Cobalt Oxide Coatings on Stainless Steel as an Electrocatalyst for Oxygen Evolution Reaction. *Electrocatalysis* **2019**, *10* (1), 63-71. DOI: 10.1007/s12678-018-0495-x.

(6) Thenuwara, A. C.; Shumlas, S. L.; Attanayake, N. H.; Cerkez, E. B.; McKendry, I. G.; Frazer, L.; Borguet, E.; Kang, Q.; Zdilla, M. J.; Sun, J.; et al. Copper-Intercalated Birnessite as a Water Oxidation Catalyst. *Langmuir* **2015**, *31* (46), 12807-12813. DOI: 10.1021/acs.langmuir.5b02936. Thenuwara, A. C.; Shumlas, S. L.; Attanayake, N. H.; Aulin, Y. V.; McKendry, I. G.; Qiao, Q.; Zhu, Y.; Borguet, E.; Zdilla, M. J.; Strongin, D. R. Intercalation of Cobalt into the Interlayer of Birnessite Improves Oxygen Evolution Catalysis. *ACS Catalysis* **2016**, *6* (11), 7739-7743. DOI: 10.1021/acscatal.6b01980.

(7) Thenuwara, A. C.; Cerkez, E. B.; Shumlas, S. L.; Attanayake, N. H.; McKendry, I. G.; Frazer, L.; Borguet, E.; Kang, Q.; Remsing, R. C.; Klein, M. L.; et al. Nickel Confined in the Interlayer Region of Birnessite: an Active Electrocatalyst for Water Oxidation. *Angew. Chem. Int. Ed.* **2016**, *55* (35), 10381-10385. DOI: 10.1002/anie.201601935.

(8) McKendry, I. G.; Kondaveeti, S. K.; Shumlas, S. L.; Strongin, D. R.; Zdilla, M. J. Decoration of the layered manganese oxide birnessite with Mn(ii/iii) gives a new water oxidation catalyst with fifty-fold turnover number enhancement. *Dalton Transactions* **2015**, *44* (29), 12981-12984, 10.1039/C5DT01436K. DOI: 10.1039/C5DT01436K.

(9) Najafpour, M. M.; Ehrenberg, T.; Wiechen, M.; Kurz, P. Calcium Manganese(III) Oxides (CaMn2O4·x H2O) as Biomimetic Oxygen-Evolving Catalysts. *Angew. Chem. Int. Ed.* **2010**, *49* (12), 2233-2237. DOI: https://doi.org/10.1002/anie.200906745. Wiechen, M.; Zaharieva, I.; Dau, H.; Kurz, P. Layered manganese oxides for water-oxidation: alkaline earth cations influence catalytic activity in a photosystem II-like fashion. *Chemical Science* **2012**, *3* (7), 2330-2339, 10.1039/C2SC20226C. DOI: 10.1039/C2SC20226C. Shevela, D.; Koroidov, S.; Najafpour, M. M.; Messinger, J.; Kurz, P. Calcium Manganese Oxides as Oxygen Evolution Catalysts: O2 Formation Pathways Indicated by 18O-Labelling Studies. *Chemistry – A European Journal* **2011**, *17* (19), 5415-5423. DOI: 10.1002/chem.201002548. Frey, C. E.; Wiechen, M.; Kurz, P. Water-oxidation catalysis by synthetic manganese oxides - systematic variations of the calcium birnessite theme. *Dalton Transactions* **2014**, *43* (11), 4370-4379, 10.1039/C3DT52604F. DOI: 10.1039/C3DT52604F.

(10) McKendry, I. G.; Thenuwara, A. C.; Shumlas, S. L.; Peng, H.; Aulin, Y. V.; Chinnam, P. R.; Borguet, E.; Strongin, D. R.; Zdilla, M. J. Systematic Doping of Cobalt into Layered Manganese Oxide Sheets Substantially Enhances Water Oxidation Catalysis. *Inorg. Chem.* **2018**, *57* (2), 557-564. DOI: 10.1021/acs.inorgchem.7b01592.

(11) McKendry, I. G.; Mohamad, L. J.; Thenuwara, A. C.; Marshall, T.; Borguet, E.; Strongin, D. R.; Zdilla, M. J. Synergistic In-Layer Cobalt Doping and Interlayer Iron Intercalation into Layered MnO2 Produces an Efficient Water Oxidation Electrocatalyst. *ACS Energy Letters* **2018**, *3* (9), 2280-2285. DOI: 10.1021/acsenergylett.8b01217.

(12) Ding, R.; Yasini, P.; Peng, H.; Perdew, J. P.; Borguet, E.; Zdilla, M. J. Reimagining the eg1 Electronic State in Oxygen Evolution Catalysis: Oxidation-State-Modulated Superlattices as a New Type of Heterostructure for Maximizing Catalysis. *Advanced Energy Materials* **2021**, *11* (41), 2101636, https://doi.org/10.1002/aenm.202101636. DOI: https://doi.org/10.1002/aenm.202101636 (acccessed 2021/11/24).

(13) Peng, H.; McKendry, I. G.; Ding, R.; Thenuwara, A. C.; Kang, Q.; Shumlas, S. L.; Strongin, D. R.; Zdilla, M. J.; Perdew, J. P. Redox properties of birnessite from a defect perspective. *Proceedings of the National Academy of Sciences* **2017**, *114* (36), 9523-9528. DOI: 10.1073/pnas.1706836114.

(14) Bhullar, R. K.; Zdilla, M. J.; Klein, M. L.; Remsing, R. C. Effect of water frustration on water oxidation catalysis in the nanoconfined interlayers of layered manganese oxides birnessite and buserite. *Journal of Materials Chemistry A* **2021**, *9* (11), 6924-6932, 10.1039/D0TA09635K. DOI: 10.1039/D0TA09635K. Remsing, R. C.; McKendry, I. G.; Strongin, D. R.; Klein, M. L.; Zdilla, M. J. Frustrated Solvation Structures Can Enhance Electron Transfer Rates. *The Journal of Physical Chemistry Letters* **2015**, *6* (23), 4804-4808. DOI: 10.1021/acs.jpclett.5b02277.

(15) Kang, Q.; Vernisse, L.; Remsing, R. C.; Thenuwara, A. C.; Shumlas, S. L.; McKendry, I. G.; Klein, M. L.; Borguet, E.; Zdilla, M. J.; Strongin, D. R. Effect of Interlayer Spacing on the Activity of Layered Manganese Oxide Bilayer Catalysts for the Oxygen Evolution Reaction. *J. Am. Chem. Soc.* **2017**, *139* (5), 1863-1870. DOI: 10.1021/jacs.6b09184.

(16) Elmaci, G.; Ozgenc, G.; Kurz, P.; Zumreoglu-Karan, B. Enhanced water oxidation performances of birnessite and magnetic birnessite nanocomposites by transition metal ion doping. *Sustainable Energy & Fuels* **2020**, *4* (6), 3157-3166. DOI: 10.1039/d0se00301h.

(17) Maitra, U.; Naidu, B. S.; Govindaraj, A.; Rao, C. N. R. Importance of trivalency and the eg1 configuration in the photocatalytic oxidation of water by Mn and Co oxides. *Proceedings of the National Academy of Sciences* **2013**, *110* (29), 11704-11707. DOI: 10.1073/pnas.1310703110. Morgan Chan, Z.; Kitchaev, D. A.; Nelson Weker, J.; Schnedermann, C.; Lim, K.; Ceder, G.; Tumas, W.; Toney, M. F.; Nocera, D. G. Electrochemical trapping of metastable Mn3+ ions for activation of MnO2 oxygen evolution catalysts. *Proceedings of the National Academy of Sciences* **2018**, *115* (23), E5261. DOI: 10.1073/pnas.1722235115. Smith, P. F.; Deibert, B. J.; Kaushik, S.; Gardner, G.; Hwang, S.; Wang, H.; Al-Sharab, J. F.; Garfunkel, E.; Fabris, L.; Li, J.; et al. Coordination Geometry and Oxidation State Requirements of Corner-Sharing MnO6 Octahedra for Water Oxidation Catalysis: An Investigation of Manganite (γ-MnOOH). *ACS Catalysis* **2016**, *6* (3), 2089-2099. DOI: 10.1021/acscatal.6b00099. Gorlin, Y.; Jaramillo, T. F. A Bifunctional Nonprecious Metal Catalyst for Oxygen Reduction and Water Oxidation. *J. Am. Chem. Soc.* **2010**, *132* (39), 13612-13614. DOI: 10.1021/ja104587v. Cady, C. W.; Gardner, G.; Maron, Z. O.; Retuerto, M.; Go, Y. B.; Segan, S.; Greenblatt, M.; Dismukes, G. C. Tuning the Electrocatalytic Water Oxidation Properties of AB2O4 Spinel Nanocrystals: A (Li, Mg, Zn) and B (Mn, Co) Site Variants of LiMn2O4. *ACS Catalysis* **2015**, *5* (6), 3403-3410. DOI: 10.1021/acscatal.5b00265. Robinson, D. M.; Go, Y. B.; Mui, M.; Gardner, G.; Zhang, Z.; Mastrogiovanni, D.; Garfunkel, E.; Li, J.; Greenblatt, M.; Dismukes, G. C. Photochemical Water Oxidation by Crystalline Polymorphs of Manganese Oxides: Structural Requirements for Catalysis. *J. Am. Chem. Soc.* **2013**, *135* (9), 3494-3501. DOI: 10.1021/ja310286h. Birkner, N.; Nayeri, S.; Pashaei, B.; Najafpour, M. M.; Casey, W. H.; Navrotsky, A. Energetic basis of catalytic activity of layered nanophase calcium manganese oxides for water oxidation. *Proceedings of the National Academy of Sciences* **2013**, *110* (22), 8801-8806. DOI: 10.1073/pnas.1306623110. Lucht, K. P.; Mendoza-Cortes, J. L. Birnessite: A Layered Manganese Oxide To Capture Sunlight for Water-Splitting Catalysis. *Journal of Physical Chemistry C* **2015**, *119* (40), 22838-22846. DOI: 10.1021/acs.jpcc.5b07860.

(18) Deshazer, H. D.; Mantia, F. L.; Wessells, C.; Huggins, R. A.; Cui, Y. Synthesis of Nanoscale Lithium-Ion Battery Cathode Materials Using a Porous Polymer Precursor Method. *J. Electrochem. Soc.* **2011**, *158* (10), A1079. DOI: 10.1149/1.3611428.

(19) Laubach, S.; Laubach, S.; Schmidt, P. C.; Ensling, D.; Schmid, S.; Jaegermann, W.; Thißen, A.; Nikolowski, K.; Ehrenberg, H. Changes in the crystal and electronic structure of LiCoO2 and LiNiO2 upon Li intercalation and de-intercalation. *PCCP* **2009**, *11* (17), 3278-3289, 10.1039/B901200A. DOI: 10.1039/B901200A.

(20) Wang; Omomo, Y.; Sakai, N.; Fukuda, K.; Nakai, I.; Ebina, Y.; Takada, K.; Watanabe, M.; Sasaki, T. Fabrication and Characterization of Multilayer Ultrathin Films of Exfoliated MnO2 Nanosheets and Polycations. *Chem. Mater.* **2003**, *15* (15), 2873-2878. DOI: 10.1021/cm034191r.

(21) Orman, H. J.; Wiseman, P. J. Cobalt(III) lithium oxide, CoLiO2: structure refinement by powder neutron diffraction. *Acta Crystallographica Section C* **1984**, *40* (1), 12-14. DOI: doi:10.1107/S0108270184002833. Dyer, L. D.; Borie, B. S., Jr.; Smith, G. P. Alkali Metal-Nickel Oxides of the Type MNiO2. *J. Am. Chem. Soc.* **1954**, *76* (6), 1499-1503. DOI: 10.1021/ja01635a012.

(22) Guidelli, R.; Compton, R. G.; Feliu, J. M.; Gileadi, E.; Lipkowski, J.; Schmickler, W.; Trasatti, S. Defining the transfer coefficient in electrochemistry: An assessment (IUPAC Technical Report). **2014**, *86* (2), 245-258. DOI: doi:10.1515/pac-2014-5026 (acccessed 2025-01-20). Blanc, N.; Rurainsky, C.; Tschulik, K. Implications of resistance and mass transport limitations on the common Tafel approach at composite catalyst thin-film electrodes. *J. Electroanal. Chem.* **2020**, *872*, 114345. DOI: https://doi.org/10.1016/j.jelechem.2020.114345. van der Heijden, O.; Park, S.; Vos, R. E.; Eggebeen, J. J. J.; Koper, M. T. M. Tafel Slope Plot as a Tool to Analyze Electrocatalytic Reactions. *ACS Energy Letters* **2024**, *9* (4), 1871-1879. DOI: 10.1021/acsenergylett.4c00266.

(23) Narayanan, H.; Viswanathan, B.; Krishnamurthy, K. R.; Nair, H. Chapter 12 - Hydrogen from photo-electrocatalytic water splitting. In *Solar Hydrogen Production*, Calise, F., D'Accadia, M. D., Santarelli, M., Lanzini, A., Ferrero, D. Eds.; Academic Press, 2019; pp 419-486.

(24) Minaev, B. F. Spin effects in activation of hydrocarbons: The role of triplet states in catalysis. *J. Mol. Catal. A: Chem.* **2001**, *171* (1), 53-72. DOI: https://doi.org/10.1016/S1381-1169(01)00103-0.

(25) Trotochaud, L.; Young, S. L.; Ranney, J. K.; Boettcher, S. W. Nickel–Iron Oxyhydroxide Oxygen-Evolution Electrocatalysts: The Role of Intentional and Incidental Iron Incorporation. *J. Am. Chem. Soc.* **2014**, *136* (18), 6744-6753. DOI: 10.1021/ja502379c.

(26) Heyd, J.; Scuseria, G. E.; Ernzerhof, M. Hybrid functionals based on a screened Coulomb potential. *The Journal of Chemical Physics* **2003**, *118* (18), 8207-8215. DOI: 10.1063/1.1564060 (acccessed 2022/09/01). Krukau, A. V.; Vydrov, O. A.; Izmaylov, A. F.; Scuseria, G. E. Influence of the exchange screening parameter on the performance of screened hybrid functionals. *The Journal of Chemical Physics* **2006**, *125* (22), 224106. DOI: 10.1063/1.2404663 (acccessed 2021/12/16).

(27) Dovesi, R.; Erba, A.; Orlando, R.; Zicovich-Wilson, C. M.; Civalleri, B.; Maschio, L.; Rérat, M.; Casassa, S.; Baima, J.; Salustro, S.; et al. Quantum-mechanical condensed matter simulations with CRYSTAL. *WIREs Computational Molecular Science* **2018**, *8* (4), e1360, https://doi.org/10.1002/wcms.1360. DOI: https://doi.org/10.1002/wcms.1360 (acccessed 2022/09/01).

(28) Heyd, J.; Scuseria, G. E. Efficient hybrid density functional calculations in solids: Assessment of the Heyd–Scuseria–Ernzerhof screened Coulomb hybrid functional. *The Journal of Chemical Physics* **2004**, *121* (3), 1187-1192. DOI: 10.1063/1.1760074 (acccessed 2022/09/01).

(29) Vilela Oliveira, D.; Laun, J.; Peintinger, M. F.; Bredow, T. BSSE-correction scheme for consistent gaussian basis sets of double- and triple-zeta valence with polarization quality for solid-state calculations. *J. Comput. Chem.* **2019**, *40* (27), 2364-2376, https://doi.org/10.1002/jcc.26013. DOI: https://doi.org/10.1002/jcc.26013 (acccessed 2022/09/01).

(30) Grimme, S.; Antony, J.; Ehrlich, S.; Krieg, H. A consistent and accurate ab initio parametrization of density functional dispersion correction (DFT-D) for the 94 elements H-Pu. *The Journal of Chemical Physics* **2010**, *132* (15), 154104. DOI: 10.1063/1.3382344.

(31) Pulay, P. Convergence acceleration of iterative sequences. the case of scf iteration. *Chem. Phys. Lett.* **1980**, *73* (2), 393-398. DOI: https://doi.org/10.1016/0009-2614(80)80396-4. Pulay, P. Improved SCF convergence acceleration. *J. Comput. Chem.* **1982**, *3* (4), 556-560, https://doi.org/10.1002/jcc.540030413. DOI: https://doi.org/10.1002/jcc.540030413 (acccessed 2022/09/01).

(32) Hinuma, Y.; Pizzi, G.; Kumagai, Y.; Oba, F.; Tanaka, I. Band structure diagram paths based on crystallography. *Computational Materials Science* **2017**, *128*, 140-184. DOI: https://doi.org/10.1016/j.commatsci.2016.10.015. Togo, A.;

Tanaka, I. A Software Library for Crystal Symmetry Search. *ArXiv* **2018**, https://arxiv.org/abs/1808.01590.

SYNOPSIS TOC

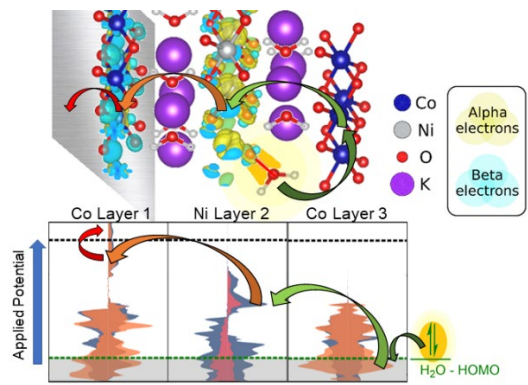

# Supporting Information for:

# Enhanced activity in layered-metal-oxide-based oxygen evolution catalysts by layer-by-layer modulation of metal ion identity.

Ran Ding,[†§1] Daniel Maldonado-Lopez, [§2] Jacob E. Henebry,[1] Jose Mendoza-Cortes,[2,3,*] Michael J. Zdilla[1,*]

[1]Deparment of Chemistry, Temple University, 1901 N. 13th St., Philadelphia, PA

[2]Department of Chemical Engineering & Materials Science, Michigan State University, East Lansing, MI 48824, USA

[3]Department of Physics & Astronomy, Michigan State University, East Lansing, Michigan 48824, USA

[§]These authors contributed equally.

Corresponding Author: * mzdilla@temple.edu, jmendoza@msu.edu

[†]Present Addresses. Department of Materials Science and Engineering, North Carolina State University, 3002 Engineering Building 1, 911 Partners Way, Raleigh NC 27695

# Contents

# Transmission Electron Microscopy

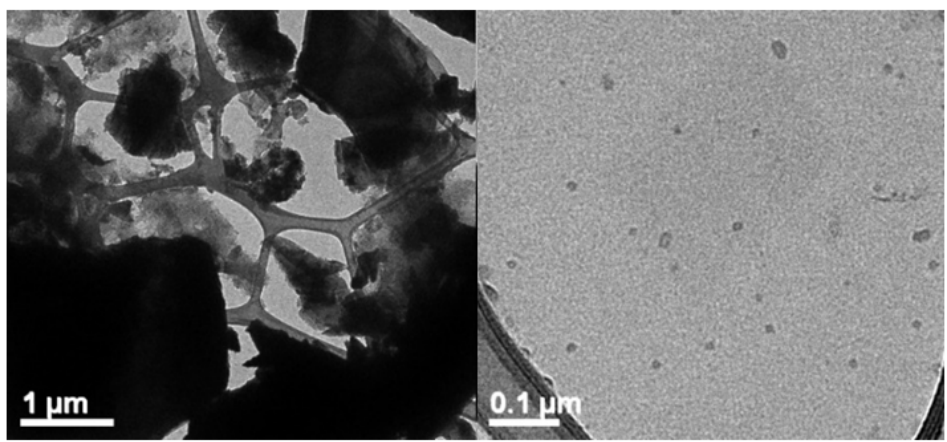

**Figure S1.** TEM images of LiCoO$_2$: Bulk (left) and single layer nanosheets (right).

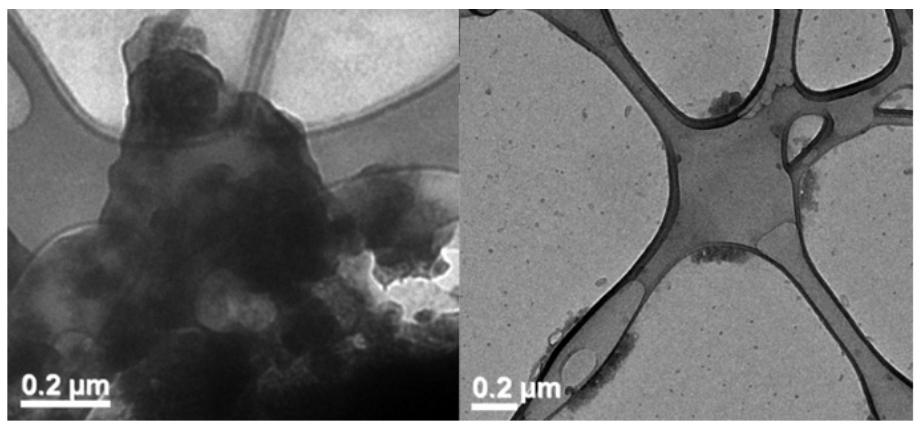

**Figure S2.** TEM images of LiNiO$_2$: Bulk (left) and single layer nanosheets (right).

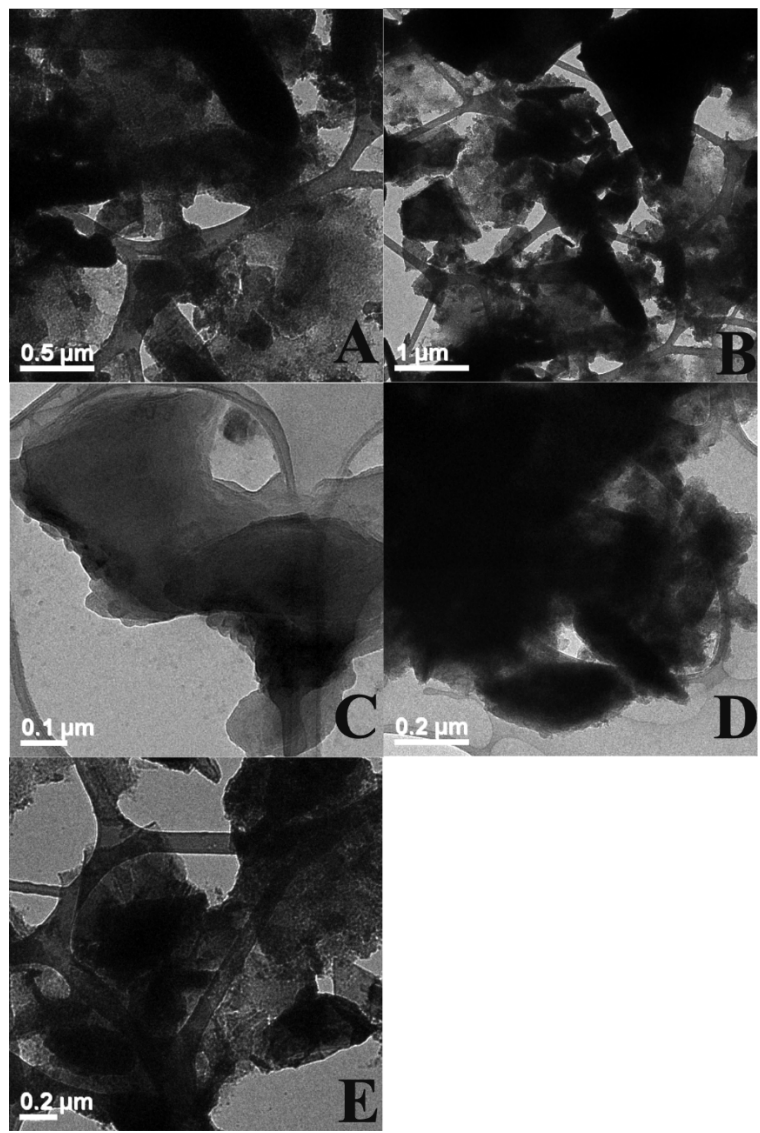

**Figure S3.** TEM images of LiCo$_x$Ni$_{(1-x)}$O$_2$. $x$ = 1 (A), 0.66 (B), 0.50 (C), 0.33 (D) and 0 (E).

## Additional Electrochemical Data

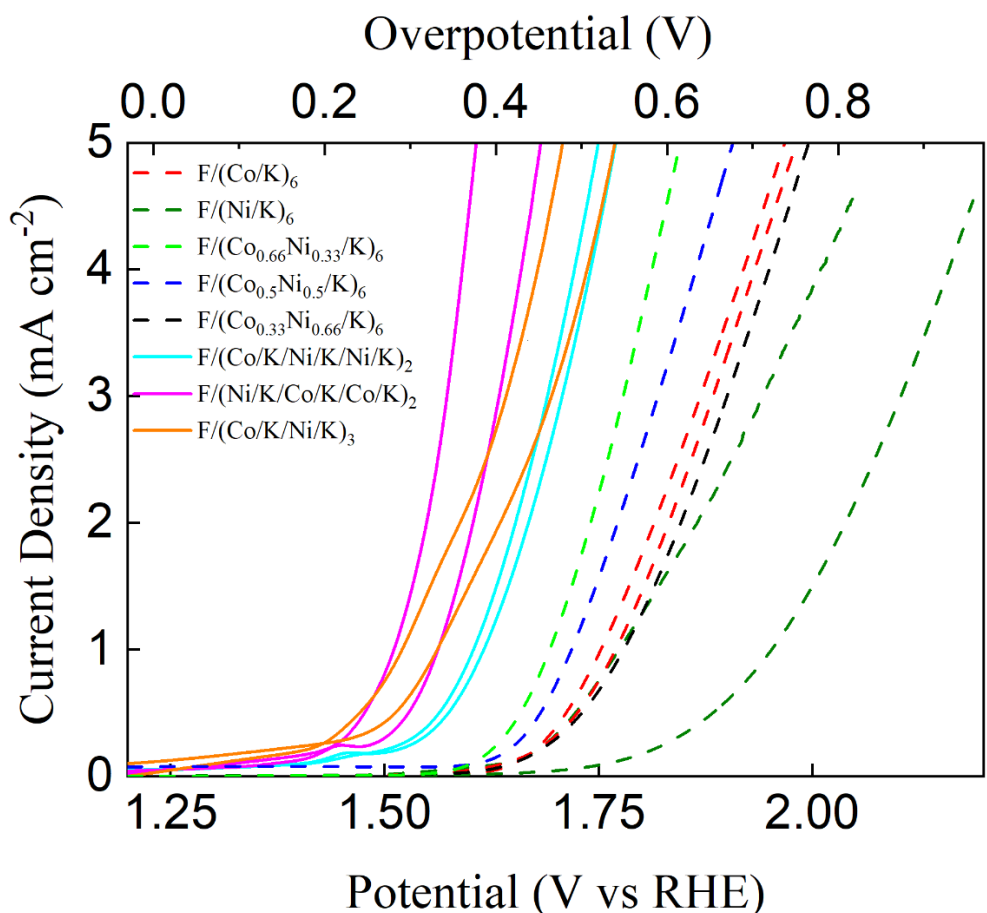

**Figure S4.** Linear sweep voltammetry of different sample batches of few-layer catalysts, illustrating slight variability (due presumably to slight differences in coverage or layer adhesion), but a robust overall trend. Each curve represents an average of three electrodes prepared from the same precursor batch. Multiple traces of the same color represent the highest- and lowest-overpotential samples we observed, so as to show the full range of catalytic activity across the samples due to differences in defect density and catalytic activity.

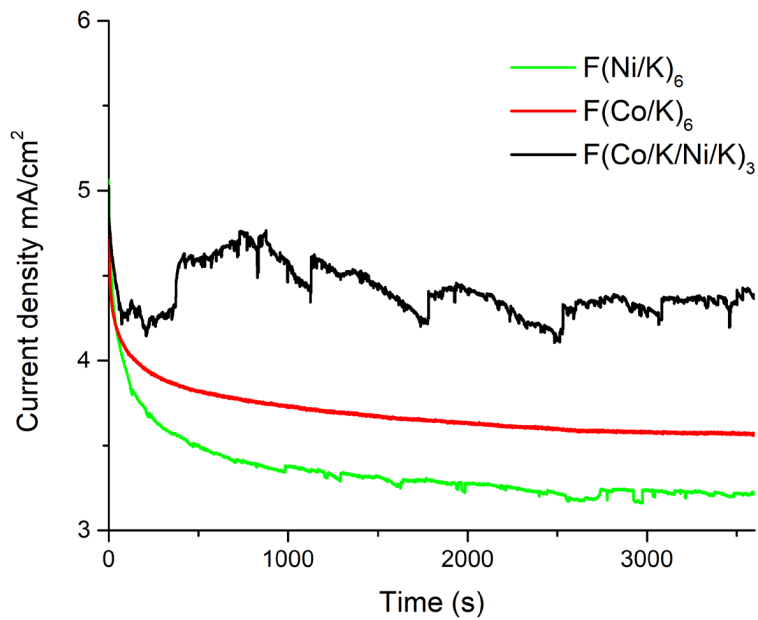

**Figure S5.** Chronoamperometry of the oxygen evolution reaction with F(Co/K)$_6$ at a potential of 1.7 V versus RHE (red); F(Ni/K)$_6$ at a potential of 1.9 V versus RHE (green); F(Co/K/Ni/K)$_3$ at a potential of 1.6 V versus RHE (black). Noise in the plots is the result of bubble formation on the electrodes.

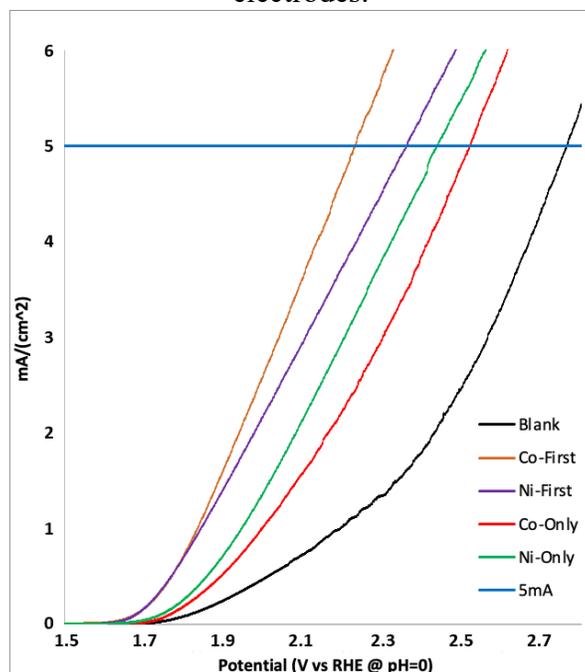

**Figure S6.** Linear sweep voltammetry of (A) F/(Ni/K)$_6$, (B) F/(Co/K)$_6$, (C) F/(Ni/K/Co/K)$_3$, and D F/(Co/K/Ni/K)$_3$. Left shows the averages of 3 samples of each catalyst, and the right shows all the LSV traces of each individual sample. For these particular catalyst batches, the KNiO$_2$ catalyst was superior to the KCoO$_2$ catalyst, but the alternating system with Co first was superior to all other arrangements.

**Table S1.** Summary of mixed cobalt and nickel layered catalysts by electrochemistry.

| Sample | Overpotential (mV) | Tafel Slope (mV/dec) |
|---|---|---|
| F/(Co/K)$_6$ | 705 | 390 |
| F/(Ni/K)$_6$ | 800 | 400 |
| F/(Co$_{0.66}$Ni$_{0.33}$/K)$_6$ | 600 | 173 |
| F/(Co$_{0.5}$Ni$_{0.5}$/K)$_6$ | 650 | 314 |
| F/(Co$_{0.33}$Ni$_{0.66}$/K)$_6$ | 740 | 245 |
| F/(Co/K/Ni/K)$_3$ | 460 | 280 |
| F/(Co/K/Ni/K/Ni/K)$_2$ | 500 | 175 |
| F/(Ni/K/Co/K/Co/K)$_2$ | 370 | 130 |

# X-ray Photoelectron Spectroscopy

To confirm controlled layer deposition using the described dipping method, we employed X-ray photoelectron spectroscopy. We observed signal for whichever layered materials were used in the dip, though signals are weak due to the atomically thin layer and the deeper escape depth of the photoelectrons in comparison to the layer thickness. We observed that when samples are alternatively dipped in $CoO_2$ nanosheets with $NiO_2$ sheets, both signals appear. Finally, postmortem XPS shows the retention of most or all of the $CoO_2$ and $NiO_2$, though some loss of intensity due to leeching is noted.

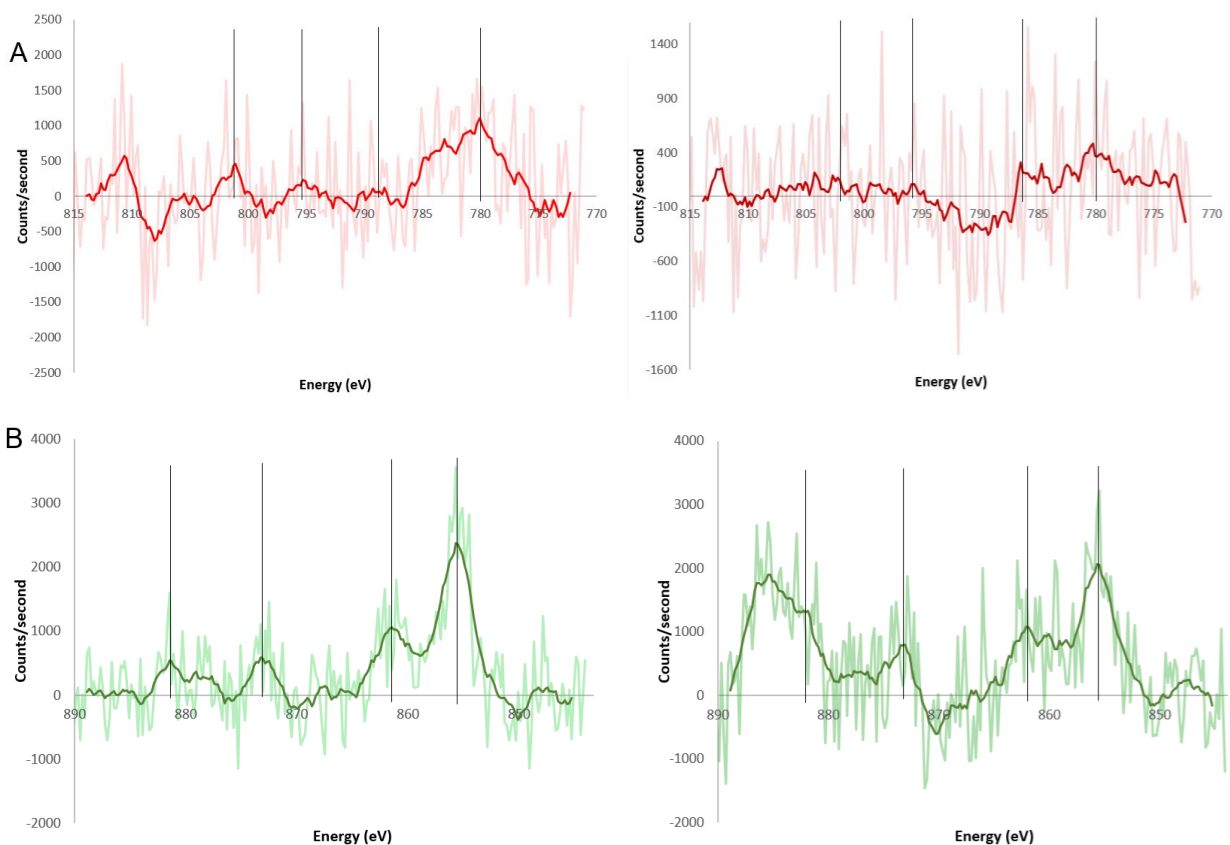

**Figure S7.** 2p XPS spectra of few-layered $KMO_2$ catalysts deposited on PEI-coated FTO. Co 2p $^3/_2$ peaks at 780eV w/ satellite 785 – 790, Co 2p ½ peaks at 796eV w/ satellite at 802 – 805. Ni 2p $^3/_2$ peaks at 855 eV w/ satellite at 865 eV, Ni 2p½ peaks at 874 eV w/ sattelite at 883 eV. (A) Co XPS of F(Co/K)$_6$ before and after catalysis. (B) Ni XPS of F(Ni/K)$_6$ before and after catalysis.

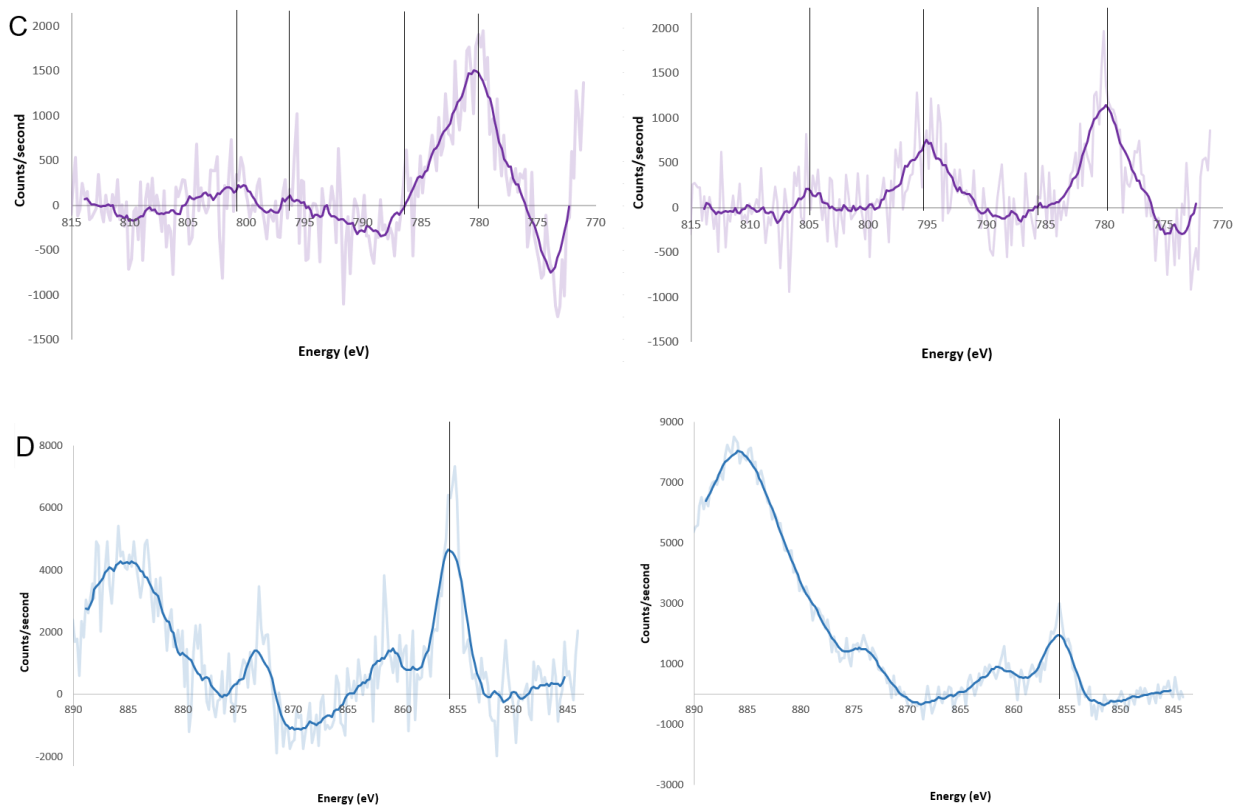

**Figure S7 (continued).** 2p XPS spectra of few-layered KMO$_2$ catalysts deposited on PEI-coated FTO. Co 2p $^3/_2$ peaks at 780eV w/ satellite 785 – 790, Co 2p ½ peaks at 796eV w/ satellite at 802 – 805. Ni 2p $^3/_2$ peaks at 855 eV w/ satellite at 865 eV, Ni 2p½ peaks at 874 eV w/ sattelite at 883 eV. (C) Co XPS of F(Co/K/Ni/K)$_3$ before and after catalysis. (D) Ni XPS of F(Co/K/Ni/K)$_3$ before and after catalysis.

# Additional Computational Results

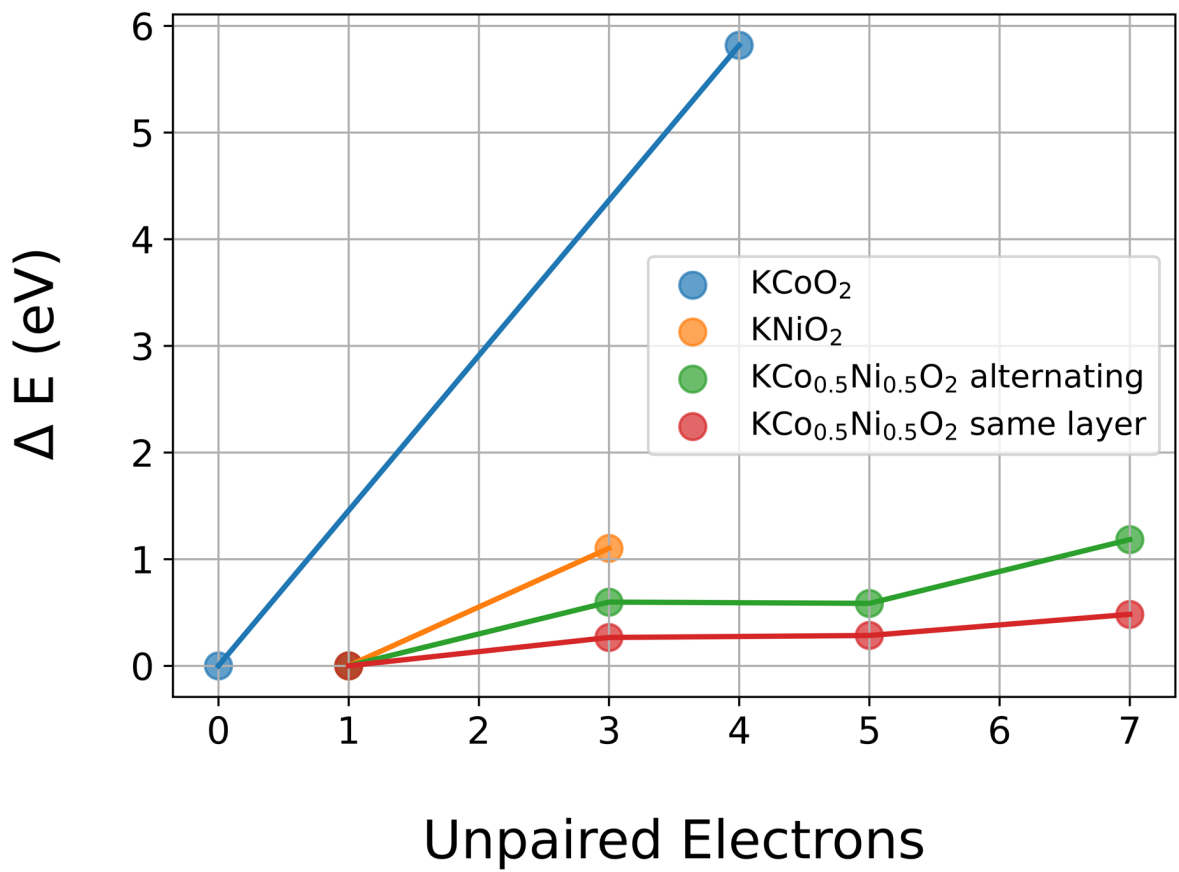

**Figure S8.** Energy comparison for different spin states. Energy difference between converged bulk structures $KCoO_2$, $KNiO_2$, alternating $KCo_{0.5}Ni_{0.5}O_2$, and same-layer $KCo_{0.5}Ni_{0.5}O_2$ at low-spin, intermediate-spin, and high-spin configurations. $KCoO_2$ and $KNiO_2$ are defined with only one transition metal in their unit cell. Therefore, they exclusively present high- and low-spin configurations.

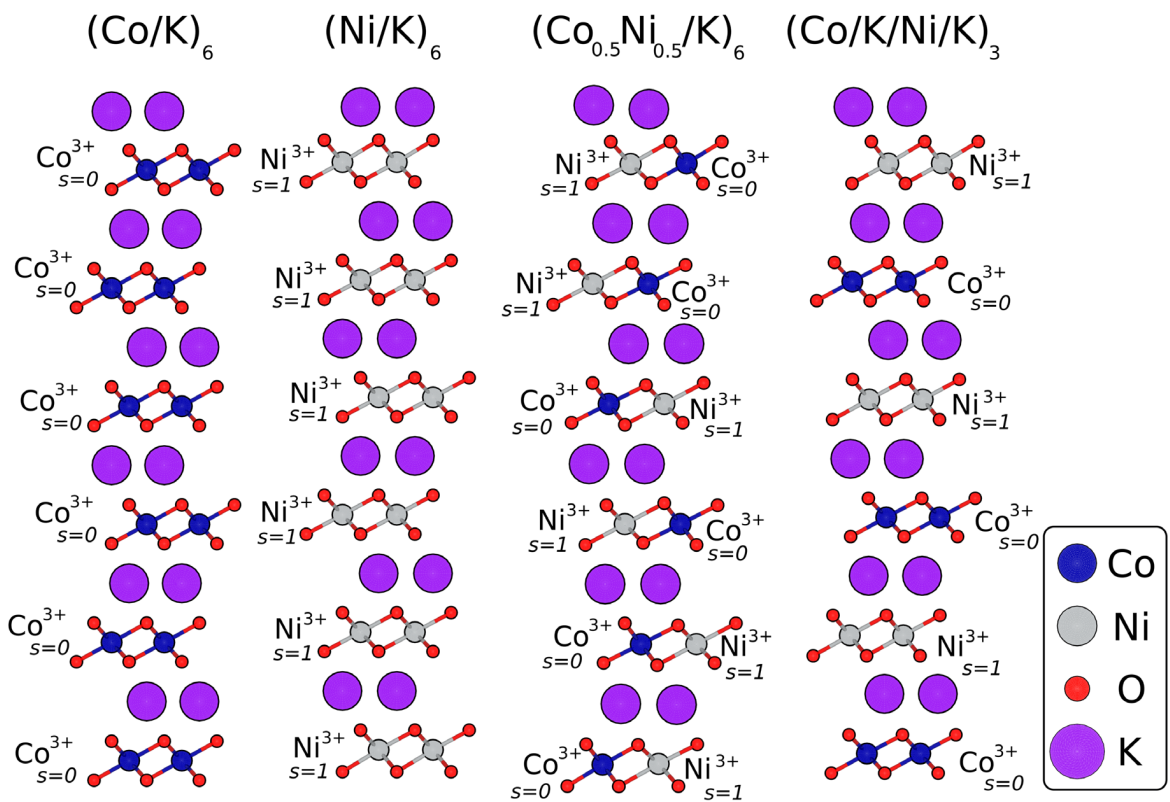

**Figure S9.** Oxidation states and unpaired electrons for Co and Ni atoms in the $(Co/K)_6$, $(Ni/K)_6$, $(Co_{0.5}Ni_{0.5}/K)_6$, and $(Co/K/Ni/K)_3$ few-layer slab structures. This information was taken from the α-β Mulliken Population Analysis of the converged models. Since oxygen and potassium ions present negligible unpaired spins, they are assumed to have a charge of 2- and 1+, respectively.

For Figure S7, it is important to point out that Mulliken Population Analysis gives a rough estimation of unpaired electrons. In general, this analysis is accurate for elements containing only *s*- and *p*-orbitals. For the $(Ni/K)_6$, $(Co_{0.5}Ni_{0.5}/K)_6$, and $(Co/K/Ni/K)_3$ slabs, the α−β spin distribution was slighly different (by ~0.2−0.5 unpaired e⁻) for outer Ni atoms, compared to inner Ni atoms. Therefore, $Ni^{2+}$ and $Ni^{4+}$ ions could be present in the structures. Magnetic moment calculations are necessary to confirm the spin structure and oxidation states.

## (Co/K)₆ CIF file

```
_cell_length_a                    2.94109216
_cell_length_b                    2.94122886
_cell_length_c                   40.00000000
_cell_angle_alpha                 90.000000
_cell_angle_beta                  90.000000
_cell_angle_gamma                120.001700
_symmetry_space_group_name_H-M    'P 1'
_symmetry_Int_Tables_number       1

loop_
_symmetry_equiv_pos_as_xyz
 'x, y, z'

loop_
  _atom_site_label
  _atom_site_type_symbol
  _atom_site_fract_x
  _atom_site_fract_y
  _atom_site_fract_z
K001   K   0.0010945638   0.0001274167   0.4296193442
O002   O   0.3336179058  -0.3341164605  -0.3867563365
O003   O  -0.3324388119   0.3333235225   0.3821493909
CO004  CO  0.3343454380  -0.3333010517   0.3579180276
CO005  CO -0.3331403391   0.3325370476  -0.3661572098
O006   O   0.0009453688  -0.0000148348   0.3342076611
O007   O   0.0004371792  -0.0006495237  -0.3413562799
K008   K   0.3333898515  -0.3341735793  -0.2932054120
K009   K  -0.3324988094   0.3331872981   0.2851768318
O010   O   0.3340008444  -0.3335678848   0.2372027708
O011   O  -0.3334609474   0.3323341291  -0.2441915144
CO012  CO  0.0006960685  -0.0002191606   0.2132572233
CO013  CO -0.0001251722  -0.0009867941  -0.2204805217
O014   O   0.3332128236  -0.3343233576  -0.1965339844
O015   O  -0.3326653461   0.3330999537   0.1895612784
K016   K   0.3338981337  -0.3336965223   0.1405911138
K017   K  -0.3332945158   0.3324807790  -0.1485594954
O018   O   0.0003993688  -0.0004810665   0.0926244132
O019   O   0.0001624304  -0.0007062869  -0.0995906472
CO020  CO  0.3335269645  -0.3340227553  -0.0758933928
CO021  CO -0.3329287863   0.3328545087   0.0686809149
```

```
O022  O   0.3337298725  -0.3338126751   0.0449836252
O023  O  -0.3331663703   0.3326272935  -0.0519483649
K024  K   0.0003182850  -0.0005949956  -0.0039828113
```

**(Ni/K)$_6$ CIF file – Note: Optimized in P3m1 symmetry (space group 156)**

```
_cell_length_a                  3.03533944
_cell_length_b                  3.03533944
_cell_length_c                 40.00000000
_cell_angle_alpha              90.000000
_cell_angle_beta               90.000000
_cell_angle_gamma             120.000000
_symmetry_space_group_name_H-M     'P 1'
_symmetry_Int_Tables_number        1

loop_
_symmetry_equiv_pos_as_xyz
 'x, y, z'

loop_
  _atom_site_label
  _atom_site_type_symbol
  _atom_site_fract_x
  _atom_site_fract_y
  _atom_site_fract_z
K001   K  -0.3333333333   0.3333333333   0.4057960009
O002   O   0.3333333333  -0.3333333333   0.3645391241
NI003  NI  0.0000000000   0.0000000000   0.3376022201
O004   O  -0.3333333333   0.3333333333   0.3131023641
K005   K   0.3333333333  -0.3333333333   0.2649300997
O006   O   0.0000000000   0.0000000000   0.2163510160
NI007  NI -0.3333333333   0.3333333333   0.1922869317
O008   O   0.3333333333  -0.3333333333   0.1681468771
K009   K   0.0000000000   0.0000000000   0.1198644760
O010   O  -0.3333333333   0.3333333333   0.0715868951
NI011  NI  0.3333333333  -0.3333333333   0.0474663059
O012   O   0.0000000000   0.0000000000   0.0232821193
K013   K  -0.3333333333   0.3333333333  -0.0249733103
O014   O   0.3333333333  -0.3333333333  -0.0732801856
NI015  NI  0.0000000000   0.0000000000  -0.0974138301
O016   O  -0.3333333333   0.3333333333  -0.1215694095
K017   K   0.3333333333  -0.3333333333  -0.1698489267
O018   O   0.0000000000   0.0000000000  -0.2181497930
NI019  NI -0.3333333333   0.3333333333  -0.2423100615
O020   O   0.3333333333  -0.3333333333  -0.2666156960
K021   K   0.0000000000   0.0000000000  -0.3144340110
```

O022  O  -0.3333333333  0.3333333333  -0.3632507796
NI023  NI  0.3333333333  -0.3333333333  -0.3871699231
 O024  O  0.0000000000  0.0000000000  -0.4086018789

**(Co/K/Ni/K)₃ CIF file**

```
_cell_length_a                2.95807100
_cell_length_b                2.95888900
_cell_length_c                40.00000000
_cell_angle_alpha              90.000000
_cell_angle_beta               90.000000
_cell_angle_gamma             118.821700
_symmetry_space_group_name_H-M     'P 1'
_symmetry_Int_Tables_number        1

loop_
_symmetry_equiv_pos_as_xyz
 'x, y, z'

loop_
  _atom_site_label
  _atom_site_type_symbol
  _atom_site_fract_x
  _atom_site_fract_y
  _atom_site_fract_z
 O001   O   0.0714980000  -0.0736930000   0.4059572500
CO002  CO   0.4056390000  -0.4080820000   0.3856705250
 O003   O  -0.2596400000   0.2574510000   0.3612481250
 K004   K   0.0801300000  -0.0815350000   0.3127136750
 O005   O   0.4106040000  -0.4123300000   0.2644902500
NI006  NI  -0.2750030000   0.2740120000   0.2397431000
 O007   O   0.0394120000  -0.0398030000   0.2150116500
 K008   K   0.3716410000  -0.3719370000   0.1668135750
 O009   O  -0.2911460000   0.2906390000   0.1185241750
CO010  CO   0.0433290000  -0.0438760000   0.0949799250
 O011   O   0.3777580000  -0.3784700000   0.0714357500
 K012   K  -0.2845180000   0.2835990000   0.0231551750
 O013   O   0.0472640000  -0.0481640000  -0.0250359500
NI014  NI   0.3619630000  -0.3616610000  -0.0497636500
 O015   O  -0.3233890000   0.3247890000  -0.0744933250
 K016   K   0.0088510000  -0.0073210000  -0.1226762500
 O017   O   0.3461230000  -0.3448230000  -0.1709703250
CO018  CO  -0.3194000000   0.3206140000  -0.1945113750
 O019   O   0.0150630000  -0.0139270000  -0.2180575750
 K020   K   0.3527270000  -0.3517130000  -0.2663490500
```

```
O021   O   -0.3140210000   0.3150210000  -0.3147987000
NI022  NI   0.0053940000  -0.0039410000  -0.3394442750
O023   O    0.3235410000  -0.3210050000  -0.3661931000
K024   K   -0.3424680000   0.3449070000  -0.4090827250
```

**(Co$_{0.5}$Ni$_{0.5}$/K)$_6$ CIF file**

```
_cell_length_a                  2.92661445
_cell_length_b                  5.25488245
_cell_length_c                 40.00000000
_cell_angle_alpha               90.000000
_cell_angle_beta                90.000000
_cell_angle_gamma               90.000000
_symmetry_space_group_name_H-M      'P 1'
_symmetry_Int_Tables_number         1

loop_
_symmetry_equiv_pos_as_xyz
 'x, y, z'

loop_
  _atom_site_label
  _atom_site_type_symbol
  _atom_site_fract_x
  _atom_site_fract_y
  _atom_site_fract_z
 K001   K   -0.5000000000  -0.0197543959   0.4094627858
 K002   K    0.0000000000   0.4769053366   0.4047455755
 O003   O   -0.5000000000  -0.3306318041   0.3653152136
 O004   O    0.0000000000   0.1343295191   0.3652307843
NI005  NI   -0.5000000000   0.3137064655   0.3383383045
CO006  CO    0.0000000000  -0.1830610813   0.3391014424
 O007   O    0.0000000000   0.4972334192   0.3150986328
 O008   O   -0.5000000000  -0.0347980723   0.3144966177
 K009   K   -0.5000000000  -0.3471390210   0.2656592720
 K010   K    0.0000000000   0.1512592557   0.2669321487
 O011   O   -0.5000000000   0.3257968264   0.2191478314
 O012   O    0.0000000000  -0.1917923569   0.2164755458
NI013  NI   -0.5000000000  -0.0264318770   0.1936905605
CO014  CO    0.0000000000   0.4728032941   0.1937933926
 O015   O    0.0000000000   0.1393546067   0.1708789761
 O016   O   -0.5000000000  -0.3811214564   0.1684927776
 K017   K   -0.5000000000   0.2963720563   0.1212524055
 K018   K    0.0000000000  -0.2037747526   0.1213325550
 O019   O   -0.5000000000  -0.0269212665   0.0741246604
 O020   O    0.0000000000   0.4535804628   0.0716656271
```

```
NI021  NI  -0.5000000000  -0.3805675084   0.0488309751
CO022  CO   0.0000000000   0.1193838570   0.0488348863
 O023   O   0.0000000000  -0.2147095508   0.0260096951
 O024   O  -0.5000000000   0.2656960495   0.0235612057
 K025   K  -0.5000000000  -0.0574467053  -0.0235910832
 K026   K  -0.0000000000   0.4425532947  -0.0235910832
 O027   O  -0.5000000000  -0.3806701159  -0.0707732497
 O028   O  -0.0000000000   0.0997721280  -0.0732157571
NI029  NI  -0.5000000000   0.2659572132  -0.0961120891
CO030  CO  -0.0000000000  -0.2340427868  -0.0961120891
 O031   O  -0.0000000000   0.4321422984  -0.1190084211
 O032   O  -0.5000000000  -0.0874154577  -0.1214509286
 K033   K  -0.5000000000  -0.4106388683  -0.1686330950
 K034   K  -0.0000000000   0.0893611317  -0.1686330950
 O035   O  -0.5000000000   0.2661377211  -0.2158152614
 O036   O  -0.0000000000  -0.2534200349  -0.2182577689
NI037  NI  -0.5000000000  -0.0872349498  -0.2411541009
CO038  CO  -0.0000000000   0.4127650502  -0.2411541009
 O039   O  -0.0000000000   0.0789501354  -0.2640504329
 O040   O  -0.5000000000  -0.4406076207  -0.2664929404
 K041   K  -0.5000000000   0.2361689687  -0.3136751068
 K042   K  -0.0000000000  -0.2638310313  -0.3136751068
 O043   O  -0.5000000000  -0.0870544419  -0.3608572732
 O044   O  -0.0000000000   0.3933878021  -0.3632997807
NI045  NI  -0.5000000000  -0.4404271128  -0.3861961127
CO046  CO  -0.0000000000   0.0595728872  -0.3861961127
 O047   O  -0.0000000000  -0.2742420276  -0.4090924447
 O048   O  -0.5000000000   0.2062002164  -0.4115349521
```